\documentclass[12pt]{article}
\usepackage{color}
\usepackage[latin1]{inputenc}
\usepackage[T1]{fontenc}
\usepackage[english]{babel}
\usepackage{amsfonts}
\usepackage{amssymb}
\usepackage{mathtools}
\usepackage{graphicx}
\usepackage{dsfont}
\usepackage{epsfig}
\usepackage{epstopdf}
\usepackage{ae}
\usepackage{amsmath}
\usepackage{amsthm}
\usepackage{thmtools}
\usepackage{endnotes}
\usepackage{setspace}
\usepackage{amsthm}
\usepackage[toc,page]{appendix}
\usepackage{float}
\usepackage{subfigure}
\usepackage{subfig}
\usepackage{multirow}
\usepackage{comment}
\usepackage{chngpage} 
\usepackage{enumerate}

\def\Prob{\mathbb{P}}
\def\E{\mathbb{E}}
\DeclareMathOperator{\VaR}{VaR}
\DeclareMathOperator{\TVaR}{TVaR}
\DeclareMathOperator{\RVaR}{RVaR}

\newtheorem{Proposition}{Proposition}[section]
\newtheorem{Theorem}{Theorem}[section]
\newtheorem{Definition}{Definition}[section]

\newtheorem{Example}{Example}[section]
\usepackage{natbib}
\usepackage{verbatim}
\usepackage{bm} 

\newcommand{\VX}{\boldsymbol{X}}
\newcommand{\Vx}{\boldsymbol{x}}

\def\melinacomment#1{\vskip 2mm\boxit{\vskip 2mm{\color{blue}\bf#1} {\color{magenta}\bf --
Melina\vskip 2mm}}\vskip 2mm}

\def\robacomment#1{\vskip 2mm\boxit{\vskip 2mm{\color{red}\bf#1} {\color{green}\bf --
Roba\vskip 2mm}}\vskip 2mm}
\providecommand{\keywords}[1]{\textbf{Keywords: } #1}
\allowdisplaybreaks
\begin{document}

\author{\textsc{Roba Bairakdar}
	\and\textsc{Lu Cao}
	\and \textsc{M\'elina Mailhot}\footnote{Department of Mathematics and Statistics, Concordia University, 1400 de Maisonneuve Blvd. West, Montr\'eal (Qu\'ebec) Canada H3G 1M8; e-mail: \texttt{melina.mailhot@concordia.ca}}
}
\title{\bf\LARGE Range Value-at-Risk: Multivariate and Extreme Values} 
\date{\today}

\maketitle

\setcounter{page}{1}

\begin{abstract}
    The concept of univariate Range Value-at-Risk, presented by \citet{cont2010robustness}, is extended in the multidimensional setting. Traditional risk measures are not well suited when dealing with heavy-tail distributions and infinite tail expectations. The multivariate definitions of robust truncated tail expectations are provided to overcome this problem. Robustness and other properties as well as  empirical estimators are derived.  Closed-form expressions and special cases in the extreme value framework are also discussed. Numerical and graphical examples are provided to examine the accuracy of the empirical estimators.
\end{abstract}


\section{Introduction}

Recent progress in understanding specific risks faced by an entity is mainly rising from the emergence of models reflecting more precisely the entity and measures that are used to quantify and represent a company's global and granular structures. Risk measures are essential for insurance companies and financial institutions for several reasons such as quantifying capital requirements to protect against unexpected future losses and to set insurance premiums for all lines of business and risk categories. Different univariate risk measures have been proposed in the literature. The most common risk measures are Value-at-Risk (VaR) and Tail Value-at-Risk (TVaR). VaR, which represents an $\alpha$-level quantile, found its way through the G-30 report, see \citet{global1993derivatives} for details. \citet{heath1999coherent} show that VaR is not a coherent risk measure in addition to not providing any information about the tail of the distribution and thus suggests other specific risk measures such as TVaR, which evaluates the average value over all VaR values at confidence levels greater than $\alpha$, which is a significant measure for heavy-tailed distributions.

Dependencies between risks needs to be taken into account to obtain accurate capital allocation and systemic risk evaluation. For example, systemic risk refers to the risks imposed by interdependencies in a system. Univariate risk measures are not suitable to be employed for heterogeneous classes of homogeneous risks. Therefore, multivariate risk measures have been developed and gained popularity in the last decade.

The notion of quantile curves is employed in \citet{embrechts2006bounds}, \citet{nappo2009kendall} and \citet{prekopa2012multivariate} to define a multivariate risk measure called upper and lower orthant VaR. Based on the same idea, \citet{cossette2013bivariate} redefine the lower and upper orthant VaR and \citet{cossette2015vector} propose the lower and upper orthant TVaR. \citet{cousin2013multivariate} develop a finite vector version of the lower and upper orthant VaR. A drawback of multivariate VaR is that it represents the boundary of the $\alpha$-level set and no additional tail information is provided, similar to the univariate VaR. Furthermore, relationships holding for univariate risk measures can be different in a multivariate setting. 

Most risk measures are defined as functions of the loss distribution which should be estimated from the data. In \citet{cont2010robustness}, risk measurement procedures are defined and analysis of robustness of different risk measures is performed. They point out the conflict between the subadditivity and robustness and propose a robust risk measure called weighted VaR (WVaR). The use of a truncated version of TVaR, defined as Range-Value-at-Risk (RVaR), is suggested by \citet{bignozzi2016parameter}. The lower and upper orthant RVaR in the multivariate setting are developed in this paper, in order to provide a new robust multivariate risk measure. We aim to study in details their properties and derive their estimators. We will also focus on extreme value distributions which can be used to model the heavy tail of the data. 

The paper is organized as follows. In Section \ref{Section:Preliminaries}, definitions and properties of the univariate $\RVaR$ are given, with examples in the Extreme Value Theory (EVT) framework. Sections \ref{Section:LowerRVaR} and \ref{Section:UpperRVaR} define the multivariate lower and upper orthant $\RVaR$, respectively. Section \ref{Section:PropertiesRVaR} presents their interesting and desirable properties, such as their behavior under transformations or translation of the multivariate variables and monotonicity. We also develop asymptotic results, the behavior with aggregate risks and we prove their robustness. In Section \ref{Section:RVaREmpirical}, we define the empirical estimator of the lower and upper orthant $\RVaR$ and we illustrate the accuracy of this estimator graphically. Concluding remarks are given in Section \ref{Section:Conclusion}

\section{Preliminaries}\label{Section:Preliminaries}
In this section, we present the univariate definition of RVaR and provide resulting measures, based on univariate RVaR, in the EVT framework and in asymptotic scenarios. 

\subsection{Univariate RVaR}
Consider a random loss variable $X$ on a probability space $(\Omega,\mathcal{F},\mathbb{P})$  with its cumulative distribution function (cdf) $F_X$. A risk measure $\rho(X)$ for a random variable (r.v.) $X$ corresponds to the required amount that has to be maintained such that the financial position $\rho(X)-X$ is acceptable. Since there are several definitions of risk measures, an appropriate choice becomes crucial for stakeholders.

\begin{Definition}
For a continuous random variable $X$ with cumulative distribution function (cdf) $F_X$, the univariate Range Value-at-Risk at significance level range $[\alpha_1, \alpha_2]\subseteq[0,1]$ is defined by
$$\RVaR_{\alpha_1,\alpha_2}(X)=\E\left[X|\VaR_{\alpha_1}(X)\leq X\leq \VaR_{\alpha_2}(X)\right]=\frac{1}{\alpha_2 -\alpha_1}\int^{\alpha_2}_{\alpha_1}\VaR_u(X)du,$$
where
$$\VaR_\alpha(X)=\inf{\{x\in\mathbb{R}:F_X(x)\geq\alpha\}}$$
is the univariate Value-at-Risk at significance level $\alpha\in[0,1]$.
\end{Definition}

For a continuous random variable $X$ with strictly increasing cdf, $\VaR_\alpha(X)=F^{-1}_X(\alpha)$, is also called the $\alpha$-quantile, where $F^{-1}_X$ is the inverse function of cdf. VaR fails to give any information beyond the level $\alpha$, However, RVaR quantifies the magnitude of the loss of the worst $100(1-\alpha_1)$ to $100(1-\alpha_2)$ cases. When $\alpha_2=1$, we obtain a special case of RVaR, which is referred to as TVaR in this article.

Robust statistics can be defined as statistics that are not unduly affected by outliers. In order to establish the robustness of RVaR, we need to define a measure of affectation. Consider a continuous random variable $X$ with cdf $F\in\mathbb{D}$ where $\mathbb{D}$ is the convex set of cdfs. Notice that a risk measure is distribution-based if $\rho(X_1)=\rho(X_2)$ when $F_{X_1}=F_{X_2}$. Hence, we use $\rho(F)\triangleq\rho(X)$ to represent the distribution-based risk measures. To quantify the sensitivity of a risk measure to the change in the distribution, we use the sensitivity function. This method is used by \citet{cont2010robustness} and can be explained as the one-sided directional derivative of the effective risk measure at $F$ in the direction $\delta_z$.

\begin{Definition}
Consider $\rho$, a distribution-based risk measure of a continuous random variable $X$ with distribution function $F\in\mathbb{D}$ where $\mathbb{D}$ is the convex set of cdfs. For $\varepsilon\in[0,1)$, set $F_\varepsilon=\varepsilon\delta_z+(1-\varepsilon)F$ such that $F_\varepsilon\in\mathbb{D}$. $\delta_z\in\mathbb{D}$ is the probability measure which gives a mass of 1 to $\{z\}$. The distribution $F_\varepsilon$ is differentiable at any $x\neq z$ and has a jump point at the point $x=z$. The sensitivity function is defined by
$$S(z)=S(z;F)\triangleq\lim_{\varepsilon\rightarrow 0^+}\frac{\rho(F_\varepsilon)-\rho(F)}{\varepsilon},$$
for any $z\in\mathbb{R}$ such that the limit exists.
\end{Definition}

The value of sensitivity function for a robust statistic will not go to infinity when $z$ becomes arbitrarily large. In other word, the bounded sensitivity function makes sure that the risk measure will not blow up when a small change happens. Accordingly, \citet{cont2010robustness} show that VaR and RVaR are robust statistics by showing that their respective sensitivity functions are bounded.

\subsection{Examples of univariate RVaR in Extreme Value Theory}\label{Section: EVT}

In this section, we will provide some examples for the discussed risk measures in the EVT framework. Most of the statistical techniques are focused on the behavior of the center of the distribution, usually the mean. However, EVT is a branch in statistics that is focused on the behavior of the tail of the distribution. There are two principle models for extreme values; the block maxima model and the peaks-over-threshold model. The \textit{block maxima} approach is used to model the largest observations from samples of identically distributed observations in successive periods. 
The \textit{peaks-over-threshold} is used to model all large observations that exceed a given high threshold value, denoted $u$. 

The limiting distribution of block maxima, from \citet{fisher1928limiting}, is given in the theorem below;

\begin{Theorem}\label{Thm: FisherTippet}
Let $X_1, \ldots, X_n$ be a sequence of independent random variables having a common distribution function $F$ and consider $M_n = \max\lbrace X_1,\ldots,X_n\rbrace$. If there exists norming constants $(a_n)$ and $(b_n)$ where $a_n\in\mathbb{R}$ and $b_n>0$ for all $n\in\mathbb{N}$ and some non-degenerate distribution function $H$ such that $$\frac{M_n-a_n}{b_n}\xrightarrow{\text{d}} H,$$ 
then $H$ is defined as the Generalized Extreme Value Distribution (GEV) given by
\begin{align*}
H_{\xi}(x)&=
\begin{dcases} 
      \exp \left\lbrace -\left(1+\xi x \right)^{-\frac{1}{\xi}}\right\rbrace,   &\ \   \xi\neq0,\\
      \exp\left\lbrace-\exp(-x)\right\rbrace,    &\ \  \xi=0,\\
  \end{dcases}
\end{align*}
where $1+\xi x>0$. A three-parameter family is obtained by defining $H_{\xi,\mu,\sigma} :=H_{\xi}\left( \frac{x-\mu }{\sigma} \right)$ for a location parameter $\mu \in \mathbb{R}$, a scale parameter $\sigma >0$, and a shape parameter $\xi \in \mathbb{R}$.

\end{Theorem}

\noindent The one-parameter GEV is the limiting distribution of the normalized maxima, but in reality, we do not know the norming constants $(a_n)$ and $(b_n)$, therefore, the three-parameter GEV provides a more general and flexible approach as it is the limiting distribution of the unnomarlized maxima.

 \citet{pickands1975statistical} and \citet{balkema1974residual} show that the theorem below  provides a very powerful result regarding the excess distribution function;
\begin{Theorem}\label{Thm:GPD}
Let $X$ be a random variable with distribution function $F$ and an upper end-point $x_F \leq \infty$. If $F$ is a distribution function that belongs to the maximum domain attraction of a GEV distribution $H_{\xi,\mu,\sigma}$, then
$$\lim_{u\to x_F}\sup_{0\leq x<x_F-u}\left| F_u(x) -G_{\xi,\sigma}(x)  \right|=0.$$
where \begin{align*}
F_u(x) = \Pr(X-u\leq x|X>u)=\frac{F(x+u)-F(u)}{1-F(u)}, \ \ \  0\leq x< x_F-u.
\end{align*}
is the excess distribution over the threshold $u$ and \begin{align*}
G_{\xi,\sigma}(x)&=
\begin{dcases} 
      1-\left(1+\xi\frac{x}{\sigma}  \right)^{-\frac{1}{\xi}},   &\ \   \xi\neq0,\\
      1-\exp\left(-\frac{x}{\sigma}  \right),    &\ \  \xi=0.\\
  \end{dcases}
\end{align*}
for $\sigma>0$, and $x\geq0$ when $\xi\geq0$, while $0\leq x\leq-\frac{\sigma}{\xi}$ when $\xi<0$. The parameters $\xi$ and $\sigma$ are referred to, respectively, as the shape and scale parameters.
\end{Theorem}
\noindent This essentially implies that $F_u \approx G_{\xi,\sigma}$ if $u$ is high enough, where $G_{\xi,\sigma}$ is called the Generalized Pareto Distribution (GPD).

\begin{Proposition}\label{Proposition:UnivariateRiskMeasuresGEV}
Assume $X \sim \text{GEV}(\mu,\sigma,\xi)$, then for $0\leq\alpha\leq1$
\begin{align*}
\VaR_{\alpha}(X)&=
\begin{dcases} 
     \mu -\frac{\sigma}{\xi}\left[1-\left(-\ln \alpha\right)^{-\xi}\right] &\ \   \xi\neq0,\\
     \E[X] -\frac{\sigma}{\xi}\left[ \Gamma(1-\xi)  -(-\ln\alpha)^{-\xi}\right]&\ \   \xi\neq0, \xi<1,\\
     \mathbb{E}[X] - \sigma\gamma - \sigma\ln(-\ln\alpha)&\ \  \xi=0,
  \end{dcases}
\end{align*}
where 
\begin{align*}
    \mathbb{E}[X] &= \begin{dcases}
         \mu + \frac{\sigma}{\xi}(\Gamma(1-\xi)-1)&\ \   \xi\neq0, \xi<1,\\
         \mu + \sigma \gamma &\ \   \xi=0,\\
         \infty &\ \   \xi\geq1,\\
    \end{dcases}
\end{align*}
where $\Gamma(x,a)$ is the incomplete Gamma function $\Gamma(x,a)=\int_a^\infty t^{x-1}e^{-t}dt$ such that $\Gamma(x)=\Gamma(x,0)$ and $\gamma$ is Euler's constant defined by $\gamma = \int_1^\infty \left(-\frac{1}{x}+\frac{1}{\lfloor x \rfloor}\right)dx$. VaR diverges for $\xi\geq1$.\\

Let $0\leq\alpha_1\leq\alpha_2\leq1$, then
\begin{align*}
\RVaR_{\alpha_1,\alpha_2}(X)=
\begin{cases} 
     \mu - \frac{\sigma}{\xi(\alpha_2-\alpha_1)}\left[(\alpha_2-\alpha_1)-\Gamma(1-\xi,-\log \alpha_2) +\Gamma(1-\xi,-\log \alpha_1) \right]&\ \   \xi\neq0,\\
     \VaR_{\alpha_i}(X) - \frac{\sigma}{\alpha_2-\alpha_1} \left[\ln(-\ln \alpha_2))-\alpha_j(\ln(-\ln \alpha_1)  - \text{li}(\alpha_2) + \text{li}(\alpha_1)\right] &\ \  \xi=0,\\
  \end{cases}
\end{align*}
for $i,j\in \lbrace 1,2\rbrace$, $i\neq j$ and where $\text{li}(x)$ is the logarithmic integral $\text{li}(x) = \int_0^x \frac{1}{\ln(t)} dt$ for $0<x<1$ and has a singularity at $x=1$.\\

As a special case, let $\alpha_2=1$, then for $\xi \neq 0$ and $\xi<1$, we have that
\begin{eqnarray*}
\TVaR_{\alpha}(X)&=& \mathbb{E}[X] -\frac{\sigma}{\xi(1-\alpha)}\left[ \Gamma(1-\xi, -\ln\alpha) -\alpha\Gamma(1-\xi)   \right]  \\
     &=& \VaR_{\alpha}(X) -\frac{\sigma}{\xi(1-\alpha)}\left[(1-\alpha)(-\ln\alpha)^{-\xi} +\Gamma(1-\xi, -\ln\alpha) - \Gamma(1-\xi) \right],
\end{eqnarray*}
and $\TVaR$ diverges for $\xi=0$ and $\xi\geq1$.\\

\end{Proposition}
\begin{proof}
See Appendix \ref{Appendix:Proof}.
\end{proof}

\noindent In addition to the risk measures, it is interesting to observe how the ratio of the risk measures behaves for large confidence levels $\alpha$.

\begin{Proposition}\label{Proposition:GEV_RVaR_limit_limit}
Assume $X\sim GEV(\mu,\sigma,\xi)$. Let $0\leq\alpha_1\leq\alpha_2\leq1$, then
\begin{align*}
   \lim_{\alpha_1 \to 1}\left[\lim_{\alpha_2 \to 1}  \frac{\RVaR_{\alpha_1,\alpha_2}(X)}{\VaR_{\alpha_1}(X)}\right]=\lim_{\alpha_1 \to 1}\left[  \frac{\TVaR_{\alpha_1}(X)}{\VaR_{\alpha_1}(X)}\right] &=\begin{dcases} 
      (1 -\xi)^{-1}  &\ \   \xi>0,\\
      1 &\ \   \xi<0.\\
  \end{dcases} 
\end{align*}
\end{Proposition}
\begin{proof}
See Appendix \ref{Appendix:Proof}.
\end{proof}

Hence, the shape parameter $\xi$ is a strong factor that affects this ratio for large values of $\alpha$. For $\xi<0$, TVaR approaches the value of VaR for high values of $\alpha$, while for $\xi>0$, TVaR becomes significantly larger than VaR.\\

\begin{Proposition}\label{Proposition:UnivariateRiskMeasuresGPD}
Consider the random variable $X$ with cdf $F$ and survival $\bar{F}$. Assume $F \in\text{MDA}(H_{\xi})$, then for $0\leq\alpha\leq1$ and $x\geq u$, where $u$ is a high threshold, we have the following
\begin{align*}
\VaR_{\alpha}(X)
   &=\begin{dcases} 
     u+\frac{\sigma}{\xi}\left(\left(\frac{1-\alpha}{\zeta_u}\right)^{-\xi} -1\right)&\ \   \xi\neq0,\\
     u-\sigma\log\left(\frac{1-\alpha}{\zeta_u}\right)  &\ \  \xi=0.\\
  \end{dcases}
\end{align*}

Let $0\leq\alpha_1\leq\alpha_2\leq1$, then for any value of $\xi$, we have that $$\RVaR_{\alpha_1,\alpha_2}(X) = \dfrac{(1-\alpha_1)\VaR_{\alpha_1}(X) - (1-\alpha_2)\VaR_{\alpha_2}(X)  }{(\alpha_2-\alpha_1)(1-\xi)}+\frac{ (\sigma-\xi u)}{(1-\xi)}.$$

As a special case, let $\alpha_2=1$, then for $\xi<1$
$$ \TVaR_{\alpha}(X)=\frac{ \VaR_\alpha(X)}{1-\xi} + \frac{  \sigma-\xi u}{1-\xi},$$
where $\zeta_u=\bar F(u)=\Prob(X>u)$, and TVaR is infinite for $\xi\geq1$.

\end{Proposition}
\begin{proof}
See Appendix \ref{Appendix:Proof}.
\end{proof}
\noindent $\RVaR$ has not been explored in the literature of Extreme Value Theory. Therefore, we have derived a closed form expression for $\RVaR$. Even though TVaR is infinite for values of $\xi>1$, RVaR exists. Thus, RVaR is useful for the cases where $\xi\geq1$, due to its ability to capture an expected value over a range of high extremes. In theory, this might not be representative of the heavy tail, however, in practice, this can be used to eliminate the issue of having an infinite mean for real data, i.e. insurance companies and financial institutions would still be interested in calculating their reserves and economic capital, and it is not possible to hold an infinite amount of reserves. In this scenario, RVaR can be used with high values of $\alpha_1$ and $\alpha_2$.\\

\noindent In addition to the results of VaR and TVaR, it is interesting to observe how the ratio of the two risk measures behaves for large confidence levels $\alpha$.

\begin{Proposition}\label{Proposition:GPD_RVaR_limit_limit}
Consider the random variable $X$ with cdf $F$. Assume $F \in\text{MDA}(H_{\xi})$, then for $0\leq\alpha\leq1$, $\xi<1$ and $x\geq u$, where $u$ is a high threshold, we have the following 
\begin{align*}
\lim_{\alpha_1 \to 1}\left[\lim_{\alpha_2 \to 1}  \frac{\RVaR_{\alpha_2,\alpha_1}(X)}{\VaR_{\alpha_1}(X)}\right]=\lim_{\alpha_1 \to 1}\left[  \frac{\TVaR_{\alpha_1}(X)}{\VaR_{\alpha_1}(X)}\right]
  &=\begin{dcases} 
      (1 -\xi)^{-1}  &\ \   \xi\geq0,\\
      1 &\ \   \xi<0.\\
  \end{dcases} 
\end{align*}
\end{Proposition}
\begin{proof}
See Appendix \ref{Appendix:Proof}.
\end{proof}

Hence, similar to the GEV distribution, the shape parameter $\xi$ is a strong factor that affects this ratio for large values of $\alpha$.\\

\section{Multivariate Lower and Upper Orthant RVaR}\label{Section:MultiRVaR}
In this section, we define the multivariate lower and upper orthant RVaR and study their properties. Examples and illustrations of the findings are provided. Finally, empirical estimators are presented.

\subsection{Lower Orthant RVaR}\label{Section:LowerRVaR}
Consider the continuous random vector $\boldsymbol{X}=(X_1,X_2,\ldots,X_d)\in \mathbb{R}_{+}^{d}$ with joint CDF $F$ and joint survival function $\bar{F}$. Define the random vector $\boldsymbol{X}_{\setminus i}=(X_1,\ldots,X_{i-1},X_{i+1},\ldots,X_d)$ with joint cdf $F_{\setminus i}$ and joint survival function $\bar{F}_{\setminus i}$, for $i=1,\ldots,d$. Let $\boldsymbol x=(x_1,\ldots,x_d)$ be a realization of $\boldsymbol X$ and consider the vector $\boldsymbol{x}_{\setminus i}=(x_1,\ldots,x_{i-1},x_{i+1},\ldots,x_d)$.

\begin{Definition}\label{Definition: LowerRVar}
Consider a continuous random vector $\VX=(X_1,X_2)$ on
the probability space $(\Omega,\mathcal{F},\mathbb{P})$ with a joint cdf $F$. The lower orthant RVaR at significance level range $[\alpha_1, \alpha_2]\subseteq[0,1]$ is given by
$$\underline{\RVaR}_{\alpha_1,\alpha_2} (\VX)= \bigcup_{i=1}^d \left\lbrace  \left(x_1,\ldots,x_{i-1},\underline{\RVaR}_{\alpha_1,\alpha_2,\boldsymbol{x}_{\setminus i}} (\VX)),x_{i+1},\ldots,x_d\right)\right\rbrace,$$
where
$$\underline{\RVaR}_{\alpha_1,\alpha_2,\boldsymbol{x}_{\setminus i}}(\VX)=\E[X_i|\underline{\VaR}_{\alpha_1,\boldsymbol{x}_{\setminus i}}(\VX)\leq X_i\leq \VaR_{\alpha_2}(X_i), \VX_{\setminus i}\leq \boldsymbol{x}_{\setminus i}],$$
for
$$\underline{\VaR}_{\alpha_1,\VaR_{\alpha_2}(\VX_{\setminus i})}(\VX)\leq x_j\leq \VaR_{\alpha_2}(X_j),\quad \text{for all } j=1,\ldots,d, i\neq j,$$
in which the lower orthant VaR at significance level $\alpha$ is defined by
$$\underline{\VaR}_{\alpha}(\VX) = \bigcup_{i=1}^d \left\lbrace  \left(x_1,\ldots,x_{i-1},\underline{\VaR}_{\alpha,\boldsymbol{x}_{\setminus i}} (\VX)),x_{i+1},\ldots,x_d\right):x_j \geq \VaR_{\alpha}(X_j),\forall j\neq i\right\rbrace,$$
where
$$\underline{\VaR}_{\alpha,\boldsymbol{x}_{\setminus i}} (\VX)=\inf\left\lbrace  x_i \in \mathbb{R}: F_{\boldsymbol{x}_{\setminus i}}(X_i)\geq \alpha \right\rbrace.$$
\end{Definition}

\begin{Proposition}\label{Proposition:RVaRLowerIntegration}
For a continuous random vector $\VX=(X_1,\ldots,X_d)$ with joint cdf $F$ and for the subvector $\VX_{\setminus i}=(X_1,\ldots,X_{i-1},X_{i+1},\ldots,X_d)$ with joint cdf $F_{\setminus i}$,  $\underline{\RVaR}_{\alpha_1,\alpha_2,\boldsymbol{x}_{\setminus i}}(\VX)$ can be restated as
$$\underline{\RVaR}_{\alpha_1,\alpha_2,\boldsymbol{x}_{\setminus i}} (\VX)=\frac{1}{F(\boldsymbol{x}_{\setminus i},\VaR_{\alpha_2}(X_i))-\alpha_1}\int_{\alpha_1}^{F(\boldsymbol{x}_{\setminus i},\VaR_{\alpha_2}(X_i))}\underline{\VaR}_{u,\boldsymbol{x}_{\setminus i}}(\VX)du,$$
for
$$\underline{\VaR}_{\alpha_1,\VaR_{\alpha_2}(\VX_{\setminus i})}(\VX)\leq x_j\leq \VaR_{\alpha_2}(X_j),\quad \text{for all } j=1,\ldots,d, \quad i\neq j.$$
\end{Proposition}
\begin{proof}
See Appendix \ref{Appendix:Proof}.
\end{proof}

\begin{figure}%
\hfill
\subfigure[]{\includegraphics[width=8cm]{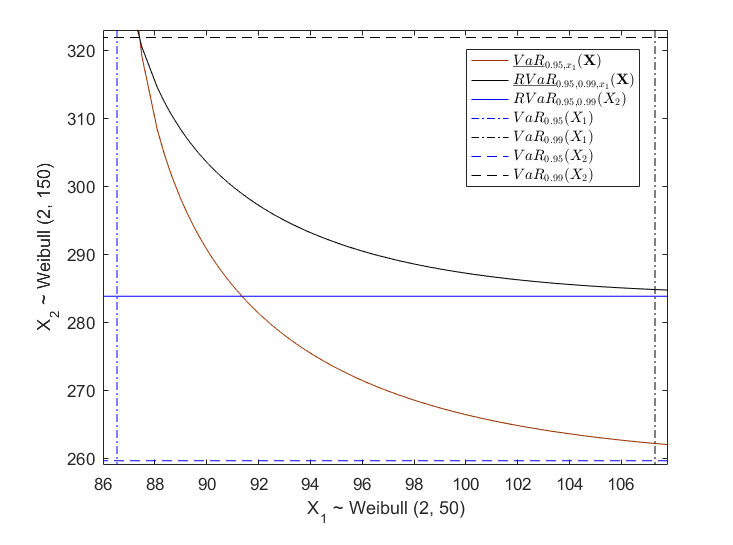}}%
\hfill
\subfigure[]{\includegraphics[width=8cm]{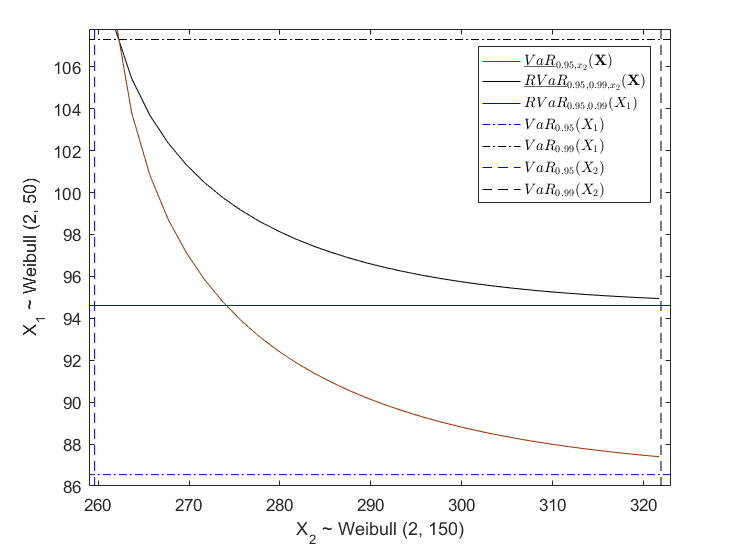}}%
\hfill
\caption{(a) Lower orthant VaR at level 0.95 and RVaR at level range $[0.95,0.99]$ for fixed values of $X_1$ and (b) Lower orthant VaR at level 0.95 and RVaR at level range $[0.95,0.99]$ for fixed values of $X_2$}%
\label{Figure5}%
\end{figure}

\begin{Example}\label{Example:RVaR_Lower}
Consider the random vector $(X_1,X_2)$ with joint cdf defined with a Gumbel copula with dependence parameter $\theta=1.5$ and marginals $X_1\sim$ Weibull (2, 50) and $X_2\sim$ Weibull (2, 150). Let the confidence level range be $\alpha_1=0.95$ and $\alpha_2=0.99$. Then, we get bivariate lower orthant $\RVaR$ in Figure \ref{Figure5}. For comparison, we plot $\underline{\VaR}_{0.95,x_i}(\VX)$ on the same graph.
\end{Example}

One can observe from Figure \ref{Figure5} that $\underline{\RVaR}_{\alpha_1,\alpha_2,x_i}(\VX)$ converges to the univariate RVaR when $x_i$ ($i=1,2$) approaches infinity. Also, when $x_i$ gets close to $\VaR_{\alpha_1}(X_i)$, $\underline{\RVaR}_{\alpha_1,\alpha_2,x_i}(\VX)$ approaches $\VaR_{\alpha_2}(X_j)$.\\

By letting $\alpha_2=1$, a special case of the lower orthant RVaR is obtained, namely the lower orthant TVaR, as defined, studied and illustrated by \citet{cossette2015vector}.

\subsection{Upper Orthant RVaR}\label{Section:UpperRVaR}

\begin{Definition}
Consider a continuous random vector $\VX=(X_1,X_2)$ on
the probability space $(\Omega,\mathcal{F},\mathbb{P})$ with a joint cdf $F$. The upper orthant $\RVaR$ at significance level range $[\alpha_1, \alpha_2]\subseteq[0,1]$ is given by
$$\overline{\RVaR}_{\alpha_1,\alpha_2} (\VX)= \bigcup_{i=1}^d \left\lbrace  \left(x_1,\ldots,x_{i-1},\overline{\RVaR}_{\alpha_1,\alpha_2,\boldsymbol{x}_{\setminus i}} (\VX)),x_{i+1},\ldots,x_d\right)\right\rbrace,$$
where
$$\overline{\RVaR}_{\alpha_1,\alpha_2,\boldsymbol{x}_{\setminus i}}(\VX)=\E[X_i|\VaR_{\alpha_1}(X_i)\leq X_i\leq \overline{\VaR}_{\alpha_2,\boldsymbol{x}_{\setminus i}}(\VX), \VX_{\setminus i}\geq \boldsymbol{x}_{\setminus i}],$$
for
$$\VaR_{\alpha_1}(X_j)\leq x_j\leq \overline{\VaR}_{\alpha_2,\VaR_{\alpha_1}(\VX_{\setminus i})}(\VX),\quad \text{for all } j=1,\ldots,d, i\neq j,$$
in which the upper orthant $\VaR$ at significance level $\alpha$ is defined by
$$\overline{\VaR}_{\alpha}(\VX) = \bigcup_{i=1}^d \left\lbrace  \left(x_1,\ldots,x_{i-1},\overline{\VaR}_{\alpha,\boldsymbol{x}_{\setminus i}} (\VX)),x_{i+1},\ldots,x_d\right):x_j \leq \VaR_{\alpha}(X_j),\forall j\neq i\right\rbrace,$$
where
$$\overline{\VaR}_{\alpha,\boldsymbol{x}_{\setminus i}} (\VX))=\inf\left\lbrace  x_i \in \mathbb{R}: \overline{F}_{\boldsymbol{x}_{\setminus i}}(X_i)\leq 1- \alpha \right\rbrace.$$
\end{Definition}

Similar to the lower orthant RVaR we can define the upper orthant RVaR
in the form of the integration of $\overline{\VaR}_{\alpha,x_i}(\VX)$.\\

\begin{Proposition}\label{Proposition:RVaRUpperIntegration}
For a continuous random vector $\VX=(X_1,\ldots,X_d)$ with joint cdf $F$ and for the subvector $\VX_{\setminus i}=(X_1,\ldots,X_{i-1},X_{i+1},\ldots,X_d)$ with joint cdf $F_{\setminus i}$,  $\overline{\RVaR}_{\alpha_1,\alpha_2,\boldsymbol{x}_{\setminus i}}(\VX)$ can be restated as
$$\overline{\RVaR}_{\alpha_1,\alpha_2,\boldsymbol{x}_{\setminus i}} (\VX)=\frac{1}{\alpha_2 - (1-\overline{F}(\boldsymbol{x}_{\setminus i},\VaR_{\alpha_1}(X_i)))}\int^{\alpha_2}_{1-\overline{F}(\boldsymbol{x}_{\setminus i},\VaR_{\alpha_1}(X_i))}\overline{\VaR}_{u,\boldsymbol{x}_{\setminus i}}(\VX)\text{d}u,$$
for
$$\VaR_{\alpha_1}(X_j)\leq x_j\leq \overline{\VaR}_{\alpha_2,\VaR_{\alpha_1}(\VX_{\setminus i})}(\VX),\quad \text{for all } j=1,\ldots,d, i\neq j,$$
\end{Proposition}
\begin{proof}
See Appendix \ref{Appendix:Proof}.
\end{proof}

\begin{figure}%
\hfill
\subfigure[]{\includegraphics[width=8cm]{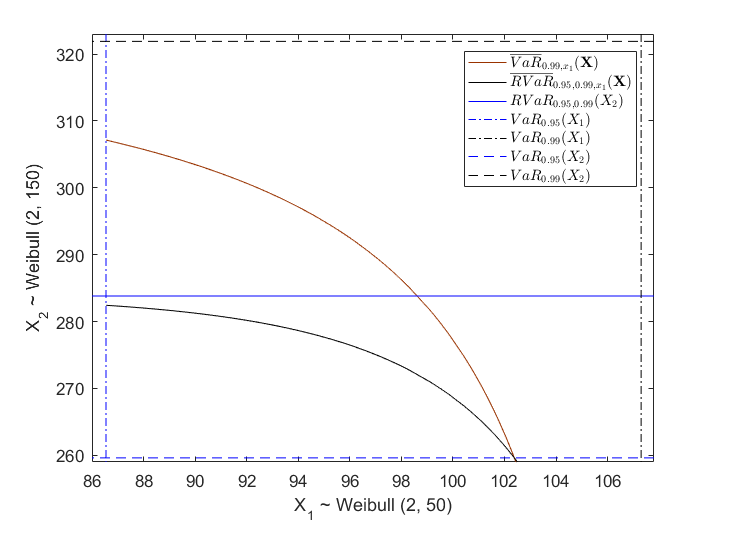}}%
\hfill
\subfigure[]{\includegraphics[width=8cm]{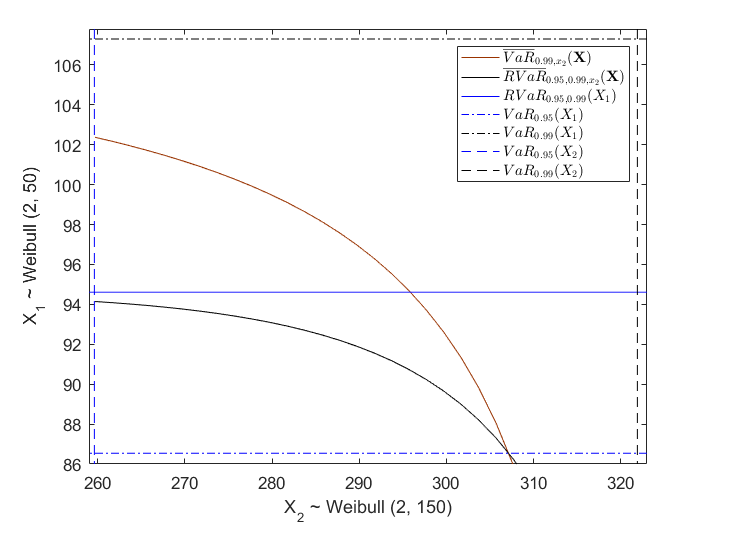}}%
\hfill
\caption{(a) Upper orthant VaR at level 0.99 and RVaR at level range $[0.95,0.99]$ for fixed values of $X_1$ and (b) Upper orthant VaR at level 0.99 and RVaR at level range $[0.95,0.99]$ for fixed values of $X_2$}%
\label{Figure6}%
\end{figure}

\begin{Example}
Consider the same random vector defined in Example~\ref{Example:RVaR_Lower}. Let the confidence level range be $\alpha_1=0.95$ and $\alpha_2=0.99$. Then, we get bivariate upper orthant $\RVaR$ in Figure \ref{Figure6}. For comparison, we plot $\overline{\VaR}_{0.99,x_i}(\VX)$ on the same graph.
\end{Example}

One observes from Figure \ref{Figure5} that $\overline{\RVaR}_{\alpha_1,\alpha_2,x_i}(\VX)$ converges to the univariate $\RVaR_{0.95,0.99}(X_j)$ when $x_i$ ($i=1,2$) gets close to the lower support of $X_i$. Also, when $x_i$ approaches $\VaR_{\alpha_2}(X_i)$, $\overline{\RVaR}_{\alpha_1,\alpha_2,x_i}(\VX)$ approaches $\VaR_{\alpha_1}(X_j)$. As a result, the curves of bivariate RVaR are bounded by the curves of univariate $\VaR$, which is similar to the univariate RVaR.\\

Analogously as for the lower orthant case, letting $\alpha_2=1$, is a special case of the upper orthant which leads to the upper orthant TVaR.

\subsection{Properties of Multivariate Lower and Upper Orthant RVaR}\label{Section:PropertiesRVaR}
For simplicity of notation and proofs, we will consider the bivariate case.
\begin{Proposition}\label{Proposition:Properties}
Let $\VX=(X_1,X_2)$ be a continuous random vector.
\begin{itemize}
\item [1.](Translation invariance) For all $\textbf{c}=(c_1,c_2)\in\mathbb{R}^2$ and $i,j=1,2$, $i\neq j$, then
$$\underline{\RVaR}_{\alpha_1,\alpha_2,x_j+c_j}(\VX+\textbf{c})=\underline{\RVaR}_{\alpha_1,\alpha_2,x_j}(\VX)+c_i,$$ $$\overline{\RVaR}_{\alpha_1,\alpha_2,x_j+c_j}(\VX+\textbf{c})=\overline{\RVaR}_{\alpha_1,\alpha_2,x_j}(\VX)+c_i.$$
\item [2.](Positive homogeneity) For all $\textbf{c}=(c_1,c_2)\in\mathbb{R}^2_{+}$ and $i,j=1,2$, $i\neq j$, then
$$\underline{\RVaR}_{\alpha_1,\alpha_2,c_jx_j}(\textbf{c}\VX)=c_i\underline{\RVaR}_{\alpha_1,\alpha_2,x_j}(\VX),$$ $$\overline{\RVaR}_{\alpha_1,\alpha_2,c_jx_j}(\textbf{c}\VX)=c_i\overline{\RVaR}_{\alpha_1,\alpha_2,x_j}(\VX).$$
\item [3.](Monotonicity) Let $\VX=(X_1,X_2)$ and $\VX'=(X'_1,X'_2)$ be two pairs of risks with joint cdf's $F_{\VX}$ and $F_{\VX'}$, respectively. If $\VX\prec_{co}\VX'$, then
$$\underline{\RVaR}_{\alpha_1,\alpha_2}(\VX')\prec\underline{\RVaR}_{\alpha_1,\alpha_2}(\VX),$$ $$\overline{\RVaR}_{\alpha_1,\alpha_2}(\VX)\prec\overline{\RVaR}_{\alpha_1,\alpha_2}(\VX').$$
\end{itemize}
\end{Proposition}
\begin{proof}
See Appendix \ref{Appendix:Proof}.
\end{proof}

Consider the random vectors $\VX_M$, $\VX_W$ and $\VX_\Pi$ which denote the monotonic, countermonotonic and independent vector, respectively. They have the following relationship
$$\VX_W\prec_{co}\VX_\Pi\prec_{co}\VX_M,$$
which means according to the Proposition \ref{Proposition:Properties}, we have
$$\underline{\RVaR}_{\alpha_1,\alpha_2}(\VX_M)\prec\underline{\RVaR}_{\alpha_1,\alpha_2}(\VX_\Pi)\prec\underline{\RVaR}_{\alpha_1,\alpha_2}(\VX_W),$$
and
$$\overline{\RVaR}_{\alpha_1,\alpha_2}(\VX_W)\prec\overline{\RVaR}_{\alpha_1,\alpha_2}(\VX_\Pi)\prec\overline{\RVaR}_{\alpha_1,\alpha_2}(\VX_M).$$

\begin{Example}
 Consider a bivariate random vector $(X_1,X_2)$ which is either comonotonic, countermotonic or independent. We obtain the lower orthant RVaR based on Proposition \ref{Proposition:RVaRLowerIntegration}. For $i,j=1,2$ ($i\neq j$),\\
\begin{align*}
&\underline{\RVaR}_{\alpha_1,\alpha_2,x_i} (\VX_\Pi)=\frac{1}{\alpha_2F_{X_i}(x_i)-\alpha_1}\int_{\alpha_1}^{\alpha_2F_{X_i}(x_i)}\VaR_{\frac{u}{F_{X_i}(x_i)}}(X_j)du,\\
&\underline{\RVaR}_{\alpha_1,\alpha_2,x_i} (\VX_M)=\frac{1}{F_{X_i}(x_i)-\alpha_1}\int^{F_{X_i}(x_i)}_{\alpha_1}\VaR_{u}(X_j)du,\\
&and\\
&\underline{\RVaR}_{\alpha_1,\alpha_2,x_i} (\VX_W)=\frac{1}{F_{X_i}(x_i)+\alpha_2-1-\alpha_1}\int_{\alpha_1}^{F_{X_i}(x_i)+\alpha_2-1}\VaR_{u-F_{X_i}(x_i)+1}(X_j)du .\end{align*}

Let the random vector above be defined with exponential marginal cdfs, i.e. $X_i\sim Exp(\lambda_i)$, then we get the following results.\\
\begin{align*}
\underline{\RVaR}_{\alpha_1,\alpha_2,x_i} (\VX_\Pi)=&\frac{1}{\alpha_2F_{X_i}(x_i)-\alpha_1}\left(\frac{1}{\lambda_i}\Bigg[(F_{X_i}(x_i)-\alpha_2F_{X_i}(x_i))\ln (1-\alpha_2)\right.\\
&\left.\left.-(F_{X_i}(x_i)-\alpha_1)\ln \left(\frac{F_{X_i}(x_i)-\alpha_1}{F_{X_i}(x_i)}\right)+(\alpha_2F_{X_i}(x_i)-\alpha_1)\right]\right),
\\
\underline{\RVaR}_{\alpha_1,\alpha_2,x_i} (\VX_M)=&\frac{1}{F_{X_i}(x_i)-\alpha_1}\left(\frac{1}{\lambda_i}\Bigg[(1-F_{X_i}(x_i))\ln (1-F_{X_i}(x_i))\right.\\
&-(1-\alpha_1)\ln \left(1-\alpha_1\right)+(F_{X_i}(x_i)-\alpha_1)\Bigg]\Bigg),
\\
\underline{\RVaR}_{\alpha_1,\alpha_2,x_i} (\VX_W)=&\frac{1}{(F_{X_i}(x_i)+\alpha_2-1)-\alpha_1}\left(\frac{1}{\lambda_i}\Bigg[(1-\alpha_2)\ln (1-\alpha_2)\right.\\
&-(F_{X_i}(x_i)-\alpha_1)\ln \left(F_{X_i}(x_i)-\alpha_1\right)+(F_{X_i}(x_i)+\alpha_2-1-\alpha_1)\Bigg]\Bigg)
.\end{align*}
\end{Example}

Now, we illustrate some examples of multivariate RVaR in the context of EVT. We present closed form expressions obtained in the independence case. Dependence between random variables is considered in section \ref{Section:RVaREmpirical}.

\begin{Example}\label{Example:GEV_Independent}
Assume $F_{X_i}\sim \text{GEV}(\mu_i,\sigma_i,\xi_i)$ and $F_{X_j}\sim \text{GEV}(\mu_j,\sigma_j,\xi_j)$. Let $0\leq\alpha\leq1$ and consider the independent copula where $C(u,v)=uv$.

Let $A=F_{X_i}(x_i), B=F(x_i,\VaR_{\alpha_2}(X_j))$ and $C=1-\bar{F}(x_i,VaR_{\alpha_1}(X_j))$, then,  \begin{align*}
\underline{\VaR}_{\alpha,x_i} (\VX)=&
    \begin{dcases}
    \mu_j-\frac{\sigma_j}{\xi_j}\left[1-\left(\ln\left(\frac{A}{\alpha}\right)\right)^{-\xi_j}\right], & \xi_i,\xi_j\neq0 \\
    \mu_j-\sigma_j\ln\left[ \ln \left(\frac{A}{\alpha}\right) \right], & \xi_i=\xi_j=0,
    \end{dcases}\\ 
\end{align*}

and
\begin{align*}
\overline{\VaR}_{\alpha,x_i} (\VX)=&
    \begin{dcases}
     \mu_j- \frac{\sigma_j}{\xi_j}\left(1-\left[\ln\left(\frac{1-A}{\alpha-A}\right)\right]^{-\xi_j} \right) , & \xi_i,\xi_j\neq0 \\
     \mu_j-\sigma_j\ln\left[\ln\left(\frac{1-A)}{\alpha-A}\right)\right] , & \xi_i=\xi_j=0.
    \end{dcases}\\ 
\end{align*}
Then by using the above results, and for $\xi_i,\xi_j \neq 0$, we obtain the multivariate lower and upper orthant RVaR, respectively represented by

\begin{align*}
\underline{\RVaR}_{\alpha_1,\alpha_2,x_i} (\VX)=& \mu_j-\frac{\sigma_j}{\xi_j}\left[1-\frac{A}{B-\alpha_1}\left[ \Gamma\left(1-\xi_j, \ln \left( \frac{A}{B}\right)\right)-\Gamma\left(1-\xi_j, \ln \left( \frac{A}{\alpha_1}\right)\right)\right]\right],\\
\overline{\RVaR}_{\alpha_1,\alpha_2,x_i} (\VX)=&\ \mu_j - \frac{\sigma_j}{\xi_j}\left[1-\frac{1-A}{\alpha_2-C} \left[\Gamma \left(1-\xi_j, \ln\left(\frac{1-A}{
\alpha_2-A} \right)   \right)   - \Gamma \left(1-\xi_j, \ln\left(\frac{1-A}{C -A}    \right)\right)\right] \right],
\end{align*}

while for $\xi_i=\xi_j=0$,
\begin{align*}
\underline{\RVaR}_{\alpha_1,\alpha_2,x_i} (\VX)=&\ \mu_j - \frac{\sigma_j}{B-\alpha_1}\left[B\ln\left(\ln\left(\frac{A}{B}\right)\right)  -\alpha_1 \ln\left( \ln\left(\frac{A}{\alpha_1}\right)\right)\right]  \\
&-\frac{\sigma_jA}{B-\alpha_1}\left[ \operatorname{Ei}\left(\ln\left(\frac{A}{\alpha_1}\right) \right)-\operatorname{Ei}\left(\ln\left(\frac{A}{B}\right)\right) \right],\\
\overline{\RVaR}_{\alpha_1,\alpha_2,x_i} (\VX)=&\ \mu_i - \frac{\sigma_j}{\alpha_2-C}\left[\left(\alpha_2-A\right)\ln\left(\ln\left(\frac{1-A}{\alpha_2-A}\right)\right)- \left(C-A\right)\ln\left(\ln\left(\frac{1-A}{C-A}\right)\right)\right]\\
&- \frac{\sigma_j\left(1-A\right)}{\alpha_2-C}\left[\operatorname{E_1}\left(\ln\left(\frac{1-A}{\alpha_2-A}\right)\right)-\operatorname{E_1}\left(\ln\left(\frac{1-A}{C-A}\right)\right)\right],
\end{align*}

where $\operatorname{Ei}(x) = -\int_{-x}^\infty\frac{e^{-t}}{t}dt$.\\

As a special case of $\RVaR$, we have that when $\alpha_2=1$,

\begin{align*}
\underline{\TVaR}_{\alpha,x_i} (\VX)=&
    \begin{dcases}
    \mu_j - \frac{\sigma_j}{\xi_j}\left[1 - \frac{ A}{A-\alpha}\left[ \Gamma\left(1-\xi_j\right)- \Gamma\left(1-\xi_j,\ln \left(\frac{A}{ \alpha}\right)\right) \right]\right], & \xi_i,\xi_j\neq0,  \\
    \infty, & \xi_i=\xi_j=0,
    \end{dcases}\\ 
\end{align*}
and
\begin{align*}
\overline{\TVaR}_{\alpha,x_i} (\VX)=&
    \begin{dcases}
     \mu_j - \frac{\sigma_j}{\xi_j}\left[1 - \dfrac{1-A}{1-\alpha}\left[\Gamma\left(1-\xi_j\right) - \Gamma\left(1-\xi_j,\ln\left(\frac{1-A}{\alpha-A}\right)\right)\right]   \right] , & \xi_i,\xi_j\neq0, \\
     \infty, & \xi_i=\xi_j=0.
    \end{dcases}\\ 
\end{align*}
\end{Example}

\begin{proof}
See Appendix \ref{Appendix:Proof}.
\end{proof}

\begin{Proposition}\label{Proposition:RVaRLimit}
Let $\VX=(X_1,X_2)$ be a pair of random variables with cdf $F_{\VX}$ and marginal distributions $F_{X_1}$ and $F_{X_2}$. Assume that $F_{\VX}$ is continuous and strictly increasing. Then, for $i,j=1,2$ and $i\neq j$,
$$\lim_{x_i\rightarrow\underline{\VaR}_{\alpha_1,\VaR_{\alpha_2}(X_j)}(\VX) }\underline{\RVaR}_{\alpha_1,\alpha_2,x_i}(\VX)=\VaR_{\alpha_2}(X_j),$$
$$\lim_{x_i\rightarrow \overline{\VaR}_{\alpha_2,\VaR_{\alpha_1}(X_j)}(\VX)}\overline{\RVaR}_{\alpha_1,\alpha_2,x_i}(\VX)=\VaR_{\alpha_1}(X_j).$$
Moreover,
$$\lim_{x_i\rightarrow u_{x_i} }\underline{\RVaR}_{\alpha_1,\alpha_2,x_i}(\VX)=\RVaR_{\alpha_1,\alpha_2}(X_j),$$
$$\lim_{x_i\rightarrow l_{x_i}}\overline{\RVaR}_{\alpha_1,\alpha_2,x_i}(\VX)=\RVaR_{\alpha_1,\alpha_2}(X_j),$$
where $u_{x_i}$ (or $l_{x_i}$) represents the upper (or lower) support of the rv $X_i$.
\end{Proposition}
\begin{proof}
See Appendix \ref{Appendix:Proof}.
\end{proof}

Now, we consider the behavior of aggregate risks defined as follows:
$$\textbf{S}=\binom{S_1}{S_2}=\sum_{i=1}^{n}\binom{X_i}{Y_i},$$
where $S_1$ and $S_2$ denote the aggregate amount of claims for two different business class respectively. $X_i$ and $Y_i$ represent the risks within each class, where $i=1,\ldots,n$, such that $S_1=\sum_{i=1}^{n}X_i$ and $S_2=\sum_{i=1}^{n}Y_i$.\\

Unlike univariate TVaR, the univariate RVaR does not satisfy the subadditivity. Hence, it seems impossible to prove that the bivariate RVaR is subadditive. However, if we suppose that $(X_1,\ldots,X_n)$ (respectively $(Y_1,\ldots,Y_n)$) is comonotonic, the following results can be obtained.\\

\begin{Proposition}\label{Proposition:Aggregate}
Let $(X_1,\ldots,X_n)$ (respectively $(Y_1,\ldots,Y_n)$) be comonotonic with cdf's $F_{X_1},\ldots,F_{X_n}$ (respectively $G_{Y_1},\ldots,G_{Y_n}$). The dependence structure between $(X_1,\ldots,X_n)$ and $(Y_1,\ldots,Y_n)$ is unknown. Then,
$$\underline{\RVaR}_{\alpha_1,\alpha_2,S_1}(\textbf{S})=\sum_{i=1}^{n}\underline{\RVaR}_{\alpha_1,\alpha_2,x_i}(X_i,Y_i),$$
$$\underline{\RVaR}_{\alpha_1,\alpha_2,S_2}(\textbf{S})=\sum_{i=1}^{n}\underline{\RVaR}_{\alpha_1,\alpha_2,y_i}(X_i,Y_i),$$
and
$$\overline{\RVaR}_{\alpha_1,\alpha_2,S_1}(\textbf{S})=\sum_{i=1}^{n}\overline{\RVaR}_{\alpha_1,\alpha_2,x_i}(X_i,Y_i),$$
$$\overline{\RVaR}_{\alpha_1,\alpha_2,S_2}(\textbf{S})=\sum_{i=1}^{n}\overline{\RVaR}_{\alpha_1,\alpha_2,y_i}(X_i,Y_i).$$
\end{Proposition}
\begin{proof}
See Appendix \ref{Appendix:Proof}.
\end{proof}

In conclusion, the bivariate RVaR has similar properties to the bivariate VaR and TVaR, such as translation invariance, positive homogeneity and monotonicity. Furthermore, it has an advantage over bivariate VaR and TVaR. Compared to bivariate VaR, bivariate TVaR and RVaR provide essential information about the tail of the distribution. Moreover, $\underline{\TVaR}_{\alpha,x_i}(\VX)$ and $\overline{\TVaR}_{\alpha,x_i}(\VX)$ will go to infinity when $X_i$ approaches $\VaR_{\alpha}(X_i)$ whereas the bivariate RVaR is bounded in the area $[\VaR_{\alpha_1}(X_i),\VaR_{\alpha_2}(X_i)]\times[\VaR_{\alpha_1}(X_j),\VaR_{\alpha_2}(X_j)]$. This measure could be useful for insurance companies that must set aside capital for risks that are sent to a reinsurer after having reached a certain level. Assume that the insurance company transfers the risks to the reinsurer when the total losses exceed VaR at level $\alpha_2$. Then, to comply to solvency capital requirements, the insurance company needs to measure the risks with truncated data. In this case, multivariate RVaR could be helpful.
\\

We will check the robustness of the estimator of bivariate RVaR. Since RVaR is distribution-based, the sensitivity function can be used to quantify the robustness.
\begin{Proposition}\label{Proposition:VaRRobust}
For a pair of continuous random variables $\VX$ with joint cdf $F(x_1,x_2)$ and marginals $F_{X_1}(x_1)$ and $F_{X_2}(x_2)$, the sensitivity function of $\underline{\VaR}_{\alpha,x_i}(\VX)$ is given by
\begin{align*}
    S(z) &=\begin{dcases} 
     -\frac{F_{X_i}(x_i)-\alpha}{f_{x_i}\left[\underline{\VaR}_{\alpha,x_i}(\VX)\right]F_{X_i}(x_i)},&\ \   z<\underline{\VaR}_{\alpha,x_i}(\VX),\\
     \frac{\alpha}{f_{x_i}\left[\underline{\VaR}_{\alpha,x_i}(\VX)\right]F_{X_i}(x_i)}, &\ \  z>\underline{\VaR}_{\alpha,x_i}(\VX),\\
      0, &\ \  \text{otherwise},\\
  \end{dcases}
\end{align*}
which is bounded. Thus, $\underline{\VaR}_{\alpha,x_i}(\VX)$ is a robust risk measure.
\end{Proposition}
\begin{proof}
See Appendix \ref{Appendix:Proof}.
\end{proof}

\begin{Proposition}\label{Proposition:RVaRRobust}
For a pair of continuous random variables $\VX$ with joint cdf $F(x_1,x_2)$ and marginals $F_{X_1}(x_1)$ and $F_{X_2}(x_2)$, let $A=F_{X_i}(x_i)$ and $B=F(x_i,\VaR_{\alpha_2}(X_j))$. Then the sensitivity function of $\underline{\RVaR}_{\alpha_1,\alpha_2,x_i}(\VX)$ is given by\\\\
$S(z)=S'(z) -\underline{\RVaR}_{\alpha_1,\alpha_2,x_i}(\VX),$\\

where

\begin{align*}
S'(z)=
\begin{dcases} 
     \frac{(A-\alpha_1)\underline{\VaR}_{\alpha_1,x_i}(\VX)-(A-B)\VaR_{\alpha_2}(X_j)}{B-\alpha_1},&\ \   z<\underline{\VaR}_{\alpha_1,x_i}(\VX),\\
     \frac{zA-\alpha_1\underline{\VaR}_{\alpha_1,x_i}(\VX)-(A-B)\VaR_{\alpha_2}(X_j)}{B-\alpha_1}, &\ \  \underline{\VaR}_{\alpha_1,x_i}(\VX)\leq z\leq \VaR_{\alpha_2}(X_j),\\
     \frac{B \VaR_{\alpha_2}(X_j)-\alpha_1\underline{\VaR}_{\alpha_1,x_i}(\VX)}{B-\alpha_1}, &\ \  z>\VaR_{\alpha_2}(X_j),\\
  \end{dcases}
\end{align*}
is a bounded function. Thus, $\underline{\RVaR}_{\alpha_1,\alpha_2,x_i}(\VX)$ is robust.
\end{Proposition}
\begin{proof}
See Appendix \ref{Appendix:Proof}.
\end{proof}

\begin{Proposition}\label{Proposition:UpperVaRRobust}
For a pair of continuous random variables $\VX$ with joint survival function $\bar{F}(x_1,x_2)$ and marginals $F_{X_1}(x_1)$ and $F_{X_2}(x_2)$, the sensitivity function of $\overline{\VaR}_{\alpha,x_i}(\VX)$ is given by
\begin{align*}
    S(z) &=\begin{dcases} 
     -\frac{1-\alpha}{f_{\bar{x}_i}\left[\overline{\VaR}_{\alpha,x_i}(\VX)\right](1-F_{X_i}(x_i))},&\ \   z<\overline{\VaR}_{\alpha,x_i}(\VX),\\
     \frac{\alpha-F_{X_i}(x_i)}{f_{\bar{x}_i}\left[\overline{\VaR}_{\alpha,x_i}(\VX)\right](1-F_{X_i}(x_i))}, &\ \  z>\overline{\VaR}_{\alpha,x_i}(\VX),\\
      0, &\ \  z=\overline{\VaR}_{\alpha,x_i}(\VX).\\
  \end{dcases}
\end{align*}
The bounded sensitivity function implies $\overline{\VaR}_{\alpha,x_i}(\VX)$ is a robust risk measure.
\end{Proposition}
\begin{proof}
See Appendix \ref{Appendix:Proof}.
\end{proof}

\begin{Proposition}\label{Proposition:RVaRUpperRobust}
For a pair of continuous random variables $\VX$ with joint survival function $\bar{F}(x_1,x_2)$ and marginals $F_{X_1}(x_1)$ and $F_{X_2}(x_2)$, let $A=F_{X_i}(x_i)$ and $C=1-\bar{F}(x_i,VaR_{\alpha_1}(X_j))$. Then the sensitivity function of $\overline{\RVaR}_{\alpha_1,\alpha_2,x_i}(\VX)$ is given by\\

$S(z)=S'(z) -\overline{\RVaR}_{\alpha_1,\alpha_2,x_i}(\VX),$\\

where

\begin{align*}
S'(z)=
\begin{dcases} 
     \frac{(1-C)\VaR_{\alpha_1}(X_j)-(1-\alpha_2)\overline{\VaR}_{\alpha_2,x_i}(\VX)}{\alpha_2-C},&\ \   z<\VaR_{\alpha_1}(X_j),\\
     \frac{z(1-A)-(C-A) \VaR_{\alpha_1}(X_j)-(1-\alpha_2)\overline{\VaR}_{\alpha_2,x_i}(\VX)}{\alpha_2-C}, &\ \  \VaR_{\alpha_1}(X_j)\leq z\leq \overline{\VaR}_{\alpha_2,x_i}(\VX),\\
      \frac{(\alpha_2-A) \overline{\VaR}_{\alpha_2,x_i}(\VX)-(C-A) \VaR_{\alpha_1}(X_j)}{\alpha_2-C}, &\ \  z>\overline{\VaR}_{\alpha_2,x_i}(\VX).\\
  \end{dcases}
\end{align*}

The bounded function proves that $\overline{\RVaR}_{\alpha_1,\alpha_2,x_i}(\VX)$ is robust.
\end{Proposition}
\begin{proof}
See Appendix \ref{Appendix:Proof}.
\end{proof}

\subsection{Empirical Estimator for Multivariate Lower and Upper Orthant RVaR}\label{Section:RVaREmpirical}

Next, we will propose empirical estimators for the lower and upper orthant RVaR, based on the estimators developed by \citet{beck2015multivariate}, and provide numerical examples.
\begin{Definition}
Consider a series of observations $\VX=(\VX_1,\ldots,\VX_d)$ with $\VX_i=(x_{1i},\ldots,x_{ni})$ and $\VX_{\setminus i}=(\VX_{1},\ldots,\VX_{i-1},\VX_{i+1},\ldots,\VX_{d})$, $i=1,\ldots,d$. Denote $F_n$ and $F_{n,\setminus i}$, the empirical cdf's (ecdf) for $\VX$ and $\VX_{\setminus i}$, $i=1,\ldots,d$, respectively. We define the estimator for the lower orthant $RVaR$ for fixed $\Vx_{\setminus i}$, $i=1,\ldots,d$, by
$$\underline{\RVaR}^n_{\alpha_1,\alpha_2,\boldsymbol{x}_{\setminus i}} (\VX)=\frac{1}{F_n(\boldsymbol{x}_{\setminus i},\VaR_{\alpha_2}(\VX_i))-\alpha_1}\int_{\alpha_1}^{F_n(\boldsymbol{x}_{\setminus i},\VaR_{\alpha_2}(\VX_i))}\underline{\VaR}^n_{u,\boldsymbol{x}_{\setminus i}}(\VX)du,$$
For $m\in\mathbb{N}$ large enough, let $s=\frac{F_n(\Vx_{ \setminus i},\VaR_{\alpha_2}(\VX_{ i}))-\alpha_1}{m}$ and $u_k=\alpha_1+ks$, then the above expression can be simplified into
\begin{align*}
\underline{\RVaR}^n_{\alpha_1,\alpha_2,\Vx_{\setminus i}} (\VX)&=\sum_{k=1}^{m}\frac{\underline{\VaR}^n_{u_k,\Vx_{\setminus i}}(\VX)\cdot s}{F_n(\Vx_{\setminus i},\VaR_{\alpha_2}(\VX_i))-\alpha_1}\\
&=\sum_{k=1}^{m}\frac{\underline{\VaR}^n_{u_k,\Vx_{\setminus i}}(\VX)}{m},
\end{align*}
where $\underline{\VaR}^n_{u,\Vx_{\setminus i}}(\VX)=\inf{\{x_i\in\mathbb{R}_{+}:F_{n,\Vx_{\setminus i}}(x_i)\geq u\}}$ is the empirical lower orthant $\VaR$ for a given $\Vx_{\setminus i}$ and $F_{n,\Vx_{\setminus i}}$ is the ecdf of $\VX$ given the same $\Vx_{\setminus i}$.
\end{Definition}
Note, $\underline{\VaR}^n_{u,\Vx_{\setminus i}}(\VX)$ is the smallest value of $\VX_i$ given $\Vx_{\setminus i}$ such that $F_n$ is larger than $u$. Similarly, we define the empirical estimator of upper orthant RVaR as follows.\\

\begin{Definition}
Consider a series of observations $\VX=(\VX_1,\ldots,\VX_d)$ with $\VX_i=(x_{1i},\ldots,x_{ni})$ and $\VX_{\setminus i}=(\VX_{1},\ldots,\VX_{i-1},\VX_{i+1},\ldots,\VX_{d})$, $i=1,\ldots,d$. Denote $\bar{F}_n$ and $\bar{F}_{n,\setminus i}$, the empirical survival functions for $\VX$ and $\VX_{\setminus i}$, $i=1,\ldots,d$, respectively. We define the estimator for the upper orthant $\RVaR$ for fixed $\Vx_{\setminus i}$, $i=1,\ldots,d$, by
$$\overline{\RVaR}^n_{\alpha_1,\alpha_2,\boldsymbol{x}_{\setminus i}} (\VX)=\frac{1}{\alpha_2 - (1-\overline{F}_n(\boldsymbol{x}_{\setminus i},\VaR_{\alpha_1}(\VX_i)))}\int^{\alpha_2}_{1-\overline{F}_n(\boldsymbol{x}_{\setminus i},\VaR_{\alpha_1}(\VX_i))}\overline{\VaR}^n_{u,\boldsymbol{x}_{\setminus i}}(\VX)\text{d}u,$$
For $m\in\mathbb{N}$ large enough. Let $s=\frac{\alpha_2-(1-\bar{F}_n(\Vx_{\setminus i},\VaR_{\alpha_1}(\VX_i)))}{m}$ and $v_k=1-\bar{F}_n(\Vx_{\setminus i},\VaR_{\alpha_1}(\VX_i))+ks$, then the above expression can be simplified into
\begin{align*}
\overline{\RVaR}^n_{\alpha_1,\alpha_2,\Vx_{\setminus i}} (\VX) & =\sum_{k=1}^{m}\frac{\overline{\VaR}^n_{v_k,\Vx_{\setminus i}}(\VX)\cdot s}{\alpha_2-(1-\bar{F}_n(\Vx_{\setminus i},\VaR_{\alpha_1}(\VX_i)))} \\
&=\sum_{k=1}^{m}\frac{\overline{\VaR}^n_{v_k,\Vx_{\setminus i}}(\VX)}{m},
\end{align*}
where $\overline{\VaR}^n_{v,\Vx_{\setminus i}}(\VX)=\inf{\{x_i\in\mathbb{R}_{+}:\bar{F}_{n,\Vx_{\setminus i}}(\Vx_i)\leq 1-v\}}$ is the empirical upper orthant $\VaR$ given $\Vx_{\setminus i}$ and $\bar{F}_{n,\Vx_{\setminus i}}$ is the empirical survival function of $\VX$ given the same $\Vx_{\setminus i}$.\\
\end{Definition}
The following proposition, based on the proof of the consistency of bivariate VaR by \citet{cousin2013multivariate}, shows the consistency of the bivariate RVaR in Hausdorff distance. For $\alpha\in(0,1)$ and $r,\zeta>0$, consider the ball

$$E=B(\{\Vx\in\mathbb{R}^2_+:|F(\Vx)-\alpha|\leq r\},\zeta).$$
Denote $m^\nabla=\inf_{\Vx\in E}\parallel(\nabla F)_{\Vx} \parallel$ as the infimum of the Euclidean norm of the gradient vector and $M_H=\sup_{\Vx\in E}\parallel(HF)_{\Vx}\parallel$ as the matrix norm of the Hessian matrix evaluated at $\Vx$ for a twice differentiable $F(x_1,x_2)$.\\

\begin{Proposition}\label{Proposition:Empirical}
Let $[\alpha_1,\alpha_2]\subset(0,1)$ and $F(x_1,x_2)$ be twice differentiable on $\mathbb{R}^2$. Assume there exists $r,\zeta>0$ such that $m^\nabla>0$ and $M_H<\infty$. Assume for each $n$, $F_n$ is continuous with probability one (wp1) and
$$\parallel F-F_n\parallel\underset{n\rightarrow\infty}{\overset{wp1}{\longrightarrow}}0.$$
Also, let $F_{n,i}$ be the consistent estimator of $F_i$. Then, we have
$$\underline{\RVaR}^n_{\alpha_1,\alpha_2,x_i}(\VX)\underset{n\longrightarrow\infty}{\overset{wp1}{\longrightarrow}}\underline{\RVaR}_{\alpha_1,\alpha_2,x_{i}}(\VX).$$
\end{Proposition}
\begin{proof}
See Appendix \ref{Appendix:Proof}.
\end{proof}

\begin{figure}%
\hfill
\subfigure[]{\includegraphics[width=8cm]{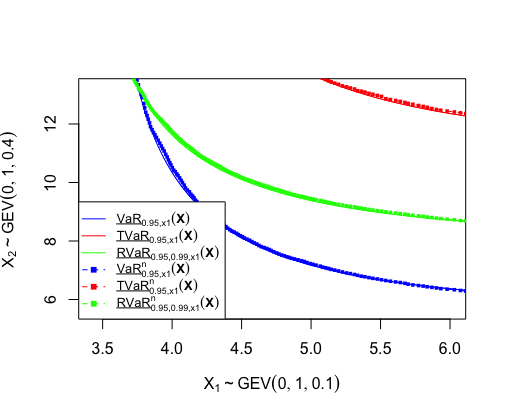}}%
\hfill
\subfigure[]{\includegraphics[width=8cm]{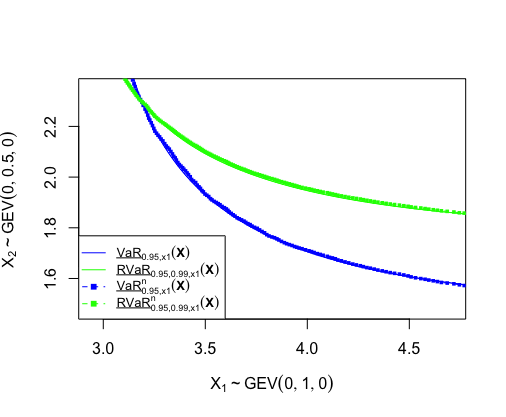}}%
\hfill
\caption{(a) Lower orthant VaR at level 0.95 and RVaR at level range $[0.95,0.99]$ for fixed values of $X_1$ for GEV samples where $\xi_1,\xi_2\neq0$ and (b) Lower orthant VaR at level 0.95 and RVaR at level range $[0.95,0.99]$ for fixed values of $X_1$ for GEV samples where $\xi_1=\xi_2=0$}
\label{Figure:GEV_simulation_Lower}%
\end{figure}

A simulation study is performed to compare the empirical estimators to the theoretical lower and upper orthant $\RVaR$. Marginally, the random variables are distributed from GEV distributions. The dependence is represented by an independent copula. 50 simulations are performed for samples of 4000 observations from each marginal distribution and for $m=250$. The results of the simulation are presented in Figure \ref{Figure:GEV_simulation_Lower}. As shown, the differeneces between the theoretical values and their empirical estimates are negligible. This could be attributed to the robustness and consistency of the empirical estimators of $\VaR$ and $\RVaR$. The accuracy of the estimates improves with the  sample size and value of $m$.

\section{Conclusion}\label{Section:Conclusion}
 In this paper, we introduce the multivariate extension of the RVaR risk measure, which results in a tool employed to assess dependent risks. This tool is particularly useful for heavy-tail distributions. Similar to its univariate counterpart, multivariate lower and upper orthant $\RVaR$ are defined as the conditional expectation of the lower and upper orthant $\VaR$ for large confidence levels. Their properties are discussed, such as translation invariance, positive homogeneity and monotonicity. Subadditivity can be satisfied for aggregated risks if each risk class is monotonic. Moreover, we develop resulting measures with specific extreme value distributions. The method of sensitivity functions to study robustness is extended for distribution-based multivariate RVaR. Finally, the empirical estimators of multivariate RVaR are proposed. The robustness and consistency of such estimators are confirmed. Furthermore, the simulations illustrate the accuracy of the empirical estimators without the need to assume any statistical distributions. RVaR may be extremely relevant in some instances where the loss distribution is characterized with an infinite mean, which results in an infinite TVaR.

\section*{Acknowledgements}
M\'elina Mailhot acknowledges financial support from the Natural Sciences and Engineering Research Council. Roba Bairakdar acknowledges financial support from the Society of Actuaries Hickman Scholar Program.
\appendix
\section{Proofs}\label{Appendix:Proof}
\subsection*{Proposition \ref{Proposition:UnivariateRiskMeasuresGEV}}
\begin{proof}
 Finding VaR is straightforward by inverting $F(\VaR_\alpha(X))=\alpha$, where $F(x)$ is the three-parameter GEV distribution.

RVaR can be derived from its definition as follows,
\begin{align*}
\RVaR_{\alpha_1,\alpha_2}(X)&=
\begin{dcases} 
     \frac{1}{\alpha_2-\alpha_1}\int_{\alpha_1}^{\alpha_2} \mu-\frac{\sigma}{\xi}\left[1-\left\lbrace -\ln w\right\rbrace^{-\xi}  \right]dw&\ \   \xi\neq0,\\
     \frac{1}{\alpha_2-\alpha_1}\int_{\alpha_1}^{\alpha_2} \mu -\sigma\ln\lbrace -\ln w \rbrace dw&\ \  \xi=0.\\
  \end{dcases}\\
  &=\begin{dcases} 
     \mu - \frac{\sigma}{\xi(\alpha_2-\alpha_1)}\left[(\alpha_2-\alpha_1)-\Gamma(1-\xi,-\ln \alpha_2) +\Gamma(1-\xi,-\ln \alpha_1) \right]&\ \   \xi\neq0,\\
     \mu - \frac{\sigma}{\alpha_2-\alpha_1} \left[\alpha_2\ln(-\ln \alpha_2)-\alpha_1\ln(-\ln \alpha_1)  - \text{li}(\alpha_2) + \text{li}(\alpha_1)\right] &\ \  \xi=0,\\
  \end{dcases}
\end{align*}

where $\text{li}(x)$ is the logarithmic integral $\text{li}(x) = \int_0^x \frac{1}{\ln(t)} dt$ for $0<x<1$ and has a singularity at $x=1$. Thus, TVaR diverges for $\xi=0$. 

TVaR can be directly derived as a special case of RVaR, when $\alpha_2=1$.

\end{proof}

\subsection*{Proposition \ref{Proposition:GEV_RVaR_limit_limit}}

\begin{proof}

When $\xi\neq0$, we have that
\begin{align*}
\lim_{\alpha_2 \to 1}  \frac{\RVaR_{\alpha_1,\alpha_2}(X)}{\VaR_{\alpha_1}(X)} &=
     \displaystyle\lim_{\alpha_2 \to 1} \dfrac{\mu - \frac{\sigma}{\xi} - \frac{\sigma}{\xi(\alpha_2-\alpha_1)}\left[-\Gamma(1-\xi,-\ln\alpha_2)+\Gamma(1-\xi,-\ln\alpha_1)    \right]}{\mu -\frac{\sigma}{\xi}\left[1-\left(-\ln \alpha_1\right)^{-\xi}\right]} \\
 &= \dfrac{\mu - \frac{\sigma}{\xi(1-\alpha_1)}\left[(1-\alpha_1)-\Gamma(1-\xi)+\Gamma(1-\xi,-\ln\alpha_1)    \right]}{\mu -\frac{\sigma}{\xi}\left[1-\left(-\ln \alpha_1\right)^{-\xi}\right]}  \\
    &= \dfrac{\TVaR_{\alpha_1}(X)}{\VaR_{\alpha_1}(X)}. 
\end{align*}

\begin{align*}
\lim_{\alpha_1 \to 1}  \frac{\TVaR_{\alpha_1}(X)}{\VaR_{\alpha_1}(X)} &=\lim_{\alpha_1 \to 1} \left[\frac{\mu - \frac{\sigma}{\xi(1-\alpha_1)} \left[(1-\alpha_1)-\Gamma(1-\xi) + \Gamma(1-\xi,-\ln\alpha_1)  \right]}{\mu -\frac{\sigma}{\xi}\left[1-\left(-\ln \alpha_1\right)^{-\xi}\right]}\right]\\
&=\begin{dcases} 
      (1 -\xi)^{-1}  &\ \   \xi>0,\\
      1 &\ \   \xi<0.\\
  \end{dcases} 
\end{align*}
\end{proof}

\subsection*{Proposition \ref{Proposition:UnivariateRiskMeasuresGPD}}

\begin{proof}
Given that $F \in\text{MDA}(H_{\xi})$, thus we can make use of Theorem \ref{Thm:GPD}, such that for a high threshold $u$, we can model $F_u$ by a GPD, where we assume that $\xi\neq0$. Note that for $x\geq u$, we have that $\Prob\left(X>x|X>u\right)=\dfrac{\Prob(X>x)}{\Prob(X>u)}=\dfrac{\bar F(x)}{\bar F(u)}$. 



Let $0\leq\alpha_1\leq\alpha_2\leq1$, then for any value of $\xi$, we have that
\begin{align*}
\RVaR_{\alpha_1,\alpha_2}(X) =& \frac{1}{\alpha_2-\alpha_1}\int_{\alpha_1}^{\alpha_2} \left[u+\frac{\sigma}{\xi}\left(\left(\frac{1-w}{\zeta_u}\right)^{-\xi} -1\right) \right]\text{d}w\\
=& \frac{1}{(\alpha_2-\alpha_1)(1-\xi)} \left[ (1-\alpha_1)\frac{\sigma}{\xi}\left( \frac{1-\alpha_1}{\zeta_u} \right)^{-\xi}  - (1-\alpha_2)\frac{\sigma}{\xi}\left( \frac{1-\alpha_2}{\zeta_u} \right)^{-\xi}   \right.\\
&+  \left.\left(u-\frac{\sigma}{\xi}\right)(\alpha_2-\alpha_1)(1-\xi)   \right].\\
=&\dfrac{(1-\alpha_1)\VaR_{\alpha_1}(X) - (1-\alpha_2)\VaR_{\alpha_2}(X)  }{(\alpha_2-\alpha_1)(1-\xi)}+\frac{ (\sigma-\xi u)}{(1-\xi)}\\
\end{align*}

\noindent If we assume that $\xi<1$, then TVaR can be obtained directly from its definition, as such
\begin{align*}
\TVaR_\alpha(X) &= \frac{1}{1-\alpha} \int_\alpha^1  \left[u+\frac{\sigma}{\xi}\left(\left(\frac{1-\omega}{\zeta_u}\right)^{-\xi} -1\right)\right] \text{d}\omega\nonumber\\
&=\frac{ \VaR_\alpha(X)}{1-\xi} + \frac{  \sigma-\xi u}{1-\xi}.
\end{align*}
Note that if $\xi>1$, the integral does not converge.\\
\end{proof}

\subsection*{Proposition \ref{Proposition:GPD_RVaR_limit_limit}}
\begin{proof}
We first observe that 
\begin{align*}
\lim_{\alpha \to 1} \VaR_{\alpha}(X) = \lim_{\alpha \to 1} \left[u+\frac{\sigma}{\xi}\left(\left(\frac{1-\alpha}{\zeta_u}\right)^{-\xi} -1\right)\right]=\begin{dcases} 
      \infty  &\ \   \xi\geq0,\\
      u-\frac{\sigma}{\xi}    &\ \  \xi<0.\\
  \end{dcases}
\end{align*}
Thus,
\begin{align*}
\lim_{\alpha_2 \to 1}  \frac{\RVaR_{\alpha_1,\alpha_2}(X)}{\VaR_{\alpha_1}(X)} &= \lim_{\alpha_2 \to 1} \left[\frac{(1-\alpha_1)}{(\alpha_2-\alpha_1)(1-\xi)} - \frac{(1-\alpha_2)\VaR_{\alpha_2}(x)}{(\alpha_2-\alpha_1)(1-\xi)\VaR_{\alpha_1}(x)}+\frac{\sigma -\xi u}{(1-\xi)\VaR_{\alpha_1}(X)}\right]\\
&=\frac{\TVaR_\alpha(X)}{\VaR_\alpha(X)}.
\end{align*}
Also,
\begin{align*}
\lim_{\alpha \to 1}  \frac{\TVaR_\alpha(X)}{\VaR_\alpha(X)}
&= \lim_{\alpha \to 1} \left[ \frac{ 1}{1-\xi} + \frac{  \sigma-\xi u}{(1-\xi)\VaR_\alpha} \right]\nonumber \\
   &=\begin{dcases} 
     (1-\xi)^{-1}&\ \   \xi\geq0,\\
     1  &\ \  \xi<0.\\
  \end{dcases}
\end{align*}

\end{proof}

\subsection*{Proposition \ref{Proposition:RVaRLowerIntegration}}
\begin{proof}
Let $F_{\setminus i}(x_i)=\Pr(X_i\leq x_i|\VX_{\setminus i}\leq \boldsymbol{x}_{\setminus i})$, then
\begin{align*}
  \underline{\RVaR}_{\alpha_1,\alpha_2,\boldsymbol{x}_{\setminus i}}(\VX)&=E[X_i|\underline{\VaR}_{\alpha_1,\boldsymbol{x}_{\setminus i}}(\VX)\leq X_i\leq \VaR_{\alpha_2}(X_i), \VX_{\setminus i}\leq \boldsymbol{x}_{\setminus i}] \\
  &=\int_{\underline{\VaR}_{\alpha_1,\boldsymbol{x}_{\setminus i}}(\VX)}^{\VaR_{\alpha_2}(X_i)}\frac{x_i \text{d}F_{\setminus i}(x_i)}{\frac{F(\boldsymbol{x}_{\setminus i},\VaR_{\alpha_2}(X_i))}{F_{\setminus i}(\boldsymbol{x}_{\setminus i})}-\frac{\alpha_1}{F_{\setminus i}(\boldsymbol{x}_{\setminus i})}} \\
  &=\frac{F_{\setminus i}(\boldsymbol{x}_{\setminus i})}{F(\boldsymbol{x}_{\setminus i},\VaR_{\alpha_2}(X_i))-\alpha_1}\int_{\underline{\VaR}_{\alpha_1,\boldsymbol{x}_{\setminus i}}(\VX)}^{\VaR_{\alpha_2}(X_i)}x_i\text{d}F_{\setminus i}(x_i)
.\end{align*}
Note that one has
$$\underline{\VaR}_{\alpha,\boldsymbol{x}_{\setminus i}}(\VX)=\VaR_{\frac{\alpha}{F_{\setminus i}(\boldsymbol{x}_{\setminus i})}}(X_i|\VX_{\setminus i}\leq\boldsymbol{x}_{\setminus i}).$$
Then, by letting $u=F_{\setminus i}(x_i)$,
\begin{align*}
\underline{\RVaR}_{\alpha_1,\alpha_2,\Vx_{\setminus i}}(\VX)
=&\frac{F_{\setminus i}(\Vx_{\setminus i})}{F(\Vx_{\setminus i},\VaR_{\alpha_2}(X_i))-\alpha_1}\int_{\alpha_1/F_{ \setminus i}(\Vx_{\setminus i})}^{F_{\setminus  i}(\VaR_{\alpha_2}(X_i))}F^{-1}_{\setminus i}(u)\text{d}u\\
=&\frac{1}{F(\Vx_{\setminus i},\VaR_{\alpha_2}(X_i))-\alpha_1}\int_{\alpha_1}^{F_{\setminus  i}(\VaR_{\alpha_2}(X_i)F_{\setminus i}(\Vx_{\setminus i}))}F^{-1}_{\setminus i}\left(\frac{u}{F_{\setminus i}(\Vx_{\setminus i})}\right)\text{d}u\\
=&\frac{1}{F(\Vx_{\setminus i},\VaR_{\alpha_2}(X_i))-\alpha_1}\int_{\alpha_1}^{F(\Vx_{\setminus i},\VaR_{\alpha_2}(X_i))}\underline{\VaR}_{u,\Vx_{\setminus i}}(\VX)\text{d}u
.\end{align*}
\\
\end{proof}

\subsection*{Proposition \ref{Proposition:RVaRUpperIntegration}}
\begin{proof}
This follows the same reasoning as for Proposition \ref{Proposition:RVaRLowerIntegration}.
\end{proof}

\subsection*{Proposition \ref{Proposition:Properties}}

\begin{proof}
We invite the reader to refer to \citet{cossette2013bivariate} for the properties of bivariate VaR to prove the results.\\

\end{proof}
\subsection*{Example \ref{Example:GEV_Independent}}
\begin{proof}
Consider the independent copula $C(u,v)=uv$ and let $A=F_{X_i}(x_i), B=F(x_i,\VaR_{\alpha_2}(X_j))$ and $C=1-\bar{F}(x_i,VaR_{\alpha_1}(X_j))$\\

Choose $u\geq F(x_1)=\alpha$. Then, for $\xi_1\neq0$, $\xi_2\neq0$, we have that
\\\\
\begin{eqnarray*}
\underline{\RVaR}_{\alpha_1,\alpha_2,x_1} (\VX)&=&\frac{1}{B-\alpha_1}\int_{\alpha_1}^{B}\underline{\VaR}_{u,x_i}(\VX)du\\
&=&\frac{1}{B-\alpha_1}\int_{\alpha_1}^{B}\mu_j-\frac{\sigma_j}{\xi_j}\left[1-\left(\ln{A}-\ln u\right)^{-\xi_j}\right]du\\
&=&\mu_j-\frac{\sigma_j}{\xi_j}\left[1-\frac{A}{B-\alpha_1}\left[ \Gamma\left(1-\xi_j, \ln \left( \frac{A}{B}\right)\right)-\Gamma\left(1-\xi_j, \ln \left( \frac{A}{\alpha_1}\right)\right)\right]\right],
\end{eqnarray*}

and for $\xi_1=\xi_2=0$, we have
\\\\
\begin{eqnarray*}
\underline{\RVaR}_{\alpha_1,\alpha_2,x_i} (\VX)&=&\frac{1}{B-\alpha_1}\int_{\alpha_1}^{B}\underline{\VaR}_{u,x_i}(\VX)du\\
&=&\frac{1}{B-\alpha_1}\int_{\alpha_1}^{B}\mu_j-\sigma_j\ln\left[\ln \left(\frac{F(x_i)}{ u}\right) \right]du\\
&=&\mu_j -\frac{\sigma_j}{B-\alpha_1}\left[u\ln\left(\ln\left(\frac{A}{u}\right)\right)-A\operatorname{Ei}\left(\ln\left(\frac{A}{u}\right)\right)\right]_{\alpha_1}^{B}\\
&&-\frac{\sigma_jA}{B-\alpha_1}\left[ \operatorname{Ei}\left(\ln\left(\frac{A}{\alpha_1}\right) \right)-\operatorname{Ei}\left(\ln\left(\frac{A}{B}\right)\right) \right],\\
\end{eqnarray*}
where $\operatorname{Ei}(x) = -\int_{-x}^\infty\frac{e^{-t}}{t}dt$.\\

Also, for $\xi_1\neq0$, $\xi_2\neq0$, we have
\begin{eqnarray*}
\underline{\TVaR}_{\alpha,x_i} (\VX)&=&\frac{1}{A-\alpha}\int_{\alpha}^{A}\mu_j-\frac{\sigma_j}{\xi_j}\left[1-\left(\ln\left(\frac{A}{\alpha}\right)\right)^{-\xi_j}\right]du\\
&=&\mu_j - \frac{\sigma_j}{\xi_j}\left[1 - \frac{ A}{A-\alpha}\left[ \Gamma\left(1-\xi_j\right)- \Gamma\left(1-\xi_j,\ln\left(\frac{ A}{ \alpha}\right)\right) \right]\right],
\end{eqnarray*}

 and for $\xi_1=\xi_2=0$, we have 
\begin{eqnarray*}
\underline{\TVaR}_{\alpha,x_i} (\VX)&=&\frac{1}{A-\alpha}\int_{\alpha}^{A}\mu_j-\sigma_j\ln\left[\ln \left(\frac{A}{ u}\right) \right]du\\
&=&\infty.
\end{eqnarray*}
An analogous reasoning applies for the result of  $\overline{\RVaR}_{\alpha_1,\alpha_2,x_i} (\VX)$ and
$\overline{\TVaR}_{\alpha,x_i} (\VX)$.

\end{proof}

\subsection*{Proposition \ref{Proposition:RVaRLimit}}
\begin{proof}
One has that
$$\lim_{x_i\rightarrow \underline{\VaR}_{\alpha_1,\VaR_{\alpha_2}(X_j)}(\VX)}\underline{\VaR}_{\alpha_1,x_i}(\VX)=\VaR_{\alpha_2}(X_j).$$ Thus, integrating this constant on the interval $[\alpha_1,F(x_i,\VaR_{\alpha_2}(X_j))]$ results in $\VaR_{\alpha_2}(X_j)$. Similarly, we can prove that when $x_i$ approaches the upper bound $\overline{\VaR}_{\alpha_2,\VaR_{\alpha_1}(X_j)}(\VX)$, $\overline{\RVaR}_{\alpha_1,\alpha_2,x_i}(\VX)$ approaches $\VaR_{\alpha_1}(X_j)$. Furthermore, we have that
$$\lim_{x_i\rightarrow u_{x_i}}\underline{\VaR}_{u,x_i}(\VX)=\VaR_{u}(X_j).$$
Combined with $F(u_{x_i},\VaR_{\alpha_2}(X_j))=\alpha_2$, we get the result that
$$\lim_{x_i\rightarrow u_{x_i} }\underline{\RVaR}_{\alpha_1,\alpha_2,x_i}(\VX)=\frac{1}{\alpha_2 -\alpha_1}\int^{\alpha_2}_{\alpha_1}\VaR_u(X_j)du=\RVaR_{\alpha_1,\alpha_2}(X_j).$$
Similarly, we can prove the limit of $\overline{\RVaR}_{\alpha_1,\alpha_2,x_i}(\VX)$ is also $\RVaR_{\alpha_1,\alpha_2}(X_j)$.
\end{proof}

\subsection*{Proposition \ref{Proposition:Aggregate}}
\begin{proof}
Define $F^{-1}_{S_1}(u)=\sum_{i=1}^{n}F^{-1}_{X_i}(u)$ and $F^{-1}_{S_2}(u)=\sum_{i=1}^{n}G^{-1}_{Y_i}(u)$. If $(X_1,\ldots,X_n)$ (respectively $(Y_1,\ldots,Y_n)$) is comonotonic, then there exists a uniform random variable $U_1$ (respectively $U_2$) such that $S_1=F^{-1}_{S_1}(U_1)$ (respectively $S_2=F^{-1}_{S_2}(U_2)$). Hence,
\begin{align*}
\underline{\RVaR}_{\alpha_1,\alpha_2,S_2}(\textbf{S})
=&\frac{\int_{\alpha_1}^{F(s_1,\VaR_{\alpha_2}(S_2))}\underline{\VaR}_{u,S_2}\left(F^{-1}_{S_1}(U_1),F^{-1}_{S_2}(U_2)\right)du}{F(s_1,\VaR_{\alpha_2}(S_2))-\alpha_1}\\
=&\frac{\int_{\alpha_1}^{F(s_1,\VaR_{\alpha_2}(S_2))}F^{-1}_{S_1}\left(\underline{\VaR}_{u,F_{S_2}(s_2)}(U_1,U_2)\right)du}{F(s_1,\VaR_{\alpha_2}(S_2))-\alpha_1}\\
=&\sum_{i=1}^{n}\frac{\int_{\alpha_1}^{F(x_i,\VaR_{\alpha_2}(Y_i))}\underline{\VaR}_{u,y_i}\left(F^{-1}_{X_i}(U_1),G^{-1}_{Y_i}(U_2)\right)du}{F(x_i,\VaR_{\alpha_2}(Y_i))-\alpha_1}\\
=&\sum_{i=1}^{n}\frac{\int_{\alpha_1}^{F(x_i,\VaR_{\alpha_2}(Y_i))}\underline{\VaR}_{u,y_i}\left(X_i,Y_i\right)du}{F(x_i,\VaR_{\alpha_2}(Y_i))-\alpha_1}\\
=&\sum_{i=1}^{n} \underline{\RVaR}_{\alpha_1,\alpha_2,y_i}(X_i,Y_i)
.\end{align*}
The other results of Proposition \ref{Proposition:Aggregate} are developed the same way.
\end{proof}

\subsection*{Proposition \ref{Proposition:VaRRobust}}

\begin{proof}
Let $F_ {x_i}(x_j)=\Pr(X_j\leq x_j|X_i\leq x_i)$ be the conditional distribution of $X_j$ knowing $X_i$, $i,j=1,2$ ($i\neq j$). For any fixed $x_i$ and $\varepsilon\in[0,1)$, set $F_{\varepsilon, x_i}(x_j)=\varepsilon\delta_{z}+(1-\varepsilon)F_{x_i}(x_j)$. The distribution $F_{\varepsilon,x_i}$ is differentiable at any $x_j\neq z$ with $F_{\varepsilon,x_i}^{'}(x_j)=(1-\varepsilon)f_{x_i}(x_j)>0$ and has a jump at the point $x_j=z$. \\
We have that
$$\underline{\VaR}_{\alpha,x_i}(\VX)=\VaR_{\frac{\alpha}{F_{X_i}(x_i)}}(X_j|X_i\leq x_i).$$
Then,
\begin{align*}
\underline{\VaR}_{\alpha,x_i}(F_{\varepsilon,x_i})
&=F^{-1}_{\varepsilon,x_i}\left(\frac{\alpha}{A}\right)\\
    &=\begin{dcases} 
     F^{-1}_{x_i}\left(\frac{\alpha}{(1-\varepsilon)F_{X_i}(x_i)}\right),&\ \   \frac{\alpha}{F_{X_i}(x_i)}<(1-\varepsilon)F_{x_i}(z),\\
     F^{-1}_{x_i}\left(\frac{\alpha/F_{X_i}(x_i)-\varepsilon}{1-\varepsilon}\right), &\ \  \frac{\alpha}{F_{X_i}(x_i)}\geq(1-\varepsilon)F_{x_i}(z)+\varepsilon,\\
      z, &\ \  \text{otherwise}.\\
  \end{dcases}
\end{align*}
As a consquence, the sensitivity function of $\underline{\VaR}_{\alpha,x_i}(\VX)$ can be evaluated by

\begin{align*}
S(z) &=\lim_{\varepsilon\rightarrow0^{+}}\frac{\underline{\VaR}_{\alpha,x_i}(F_{\varepsilon,x_i})-\underline{\VaR}_{\alpha,x_i}(F_{x_i})}{\varepsilon}\\
&=\left[\frac{d}{d\varepsilon}\underline{\VaR}_{\alpha,x_i}(F_{\varepsilon,x_i})\right]_{\varepsilon=0}\\
    &=\begin{dcases} 
     -\frac{F_{X_i}(x_i)-\alpha}{f_{x_i}\left[\underline{\VaR}_{\alpha,x_i}(\VX)\right]F_{X_i}(x_i)},&\ \   z<\underline{\VaR}_{\alpha,x_i}(\VX),\\
     \frac{\alpha}{f_{x_i}\left[\underline{\VaR}_{\alpha,x_i}(\VX)\right]F_{X_i}(x_i)}, &\ \  z>\underline{\VaR}_{\alpha,x_i}(\VX),\\
     0, &\ \  z=\underline{\VaR}_{\alpha,x_i}(\VX).\\
  \end{dcases}
\end{align*}

The result shows that $\underline{\VaR}_{\alpha,x_i}(\VX)$ has a bounded sensitivity function for any fixed $x_i$, which means it is a robust statistic. Note that this conclusion coincides with the one associated with the univariate VaR.
\end{proof}

\subsection*{Proposition \ref{Proposition:RVaRRobust}}

\begin{proof}
 Let $A=F_{X_i}(x_i)$ and $B=F(x_i,\VaR_{\alpha_2}(X_j))$. Then the sensitivity function of $$\underline{\RVaR}_{\alpha_1,\alpha_2,x_i}(\VX)=\frac{1}{B-\alpha_1}\int_{\alpha_1}^{B}\underline{\VaR}_{u,x_i}(\VX)du,$$ is given by
\begin{align*}
S(z)=&\frac{1}{B-\alpha_1}\int_{\alpha_1}^{B}\left[\lim_{\varepsilon\rightarrow0^{+}}\frac{\underline{\VaR}_{u,x_i}(F_{\varepsilon,x_i})-\underline{\VaR}_{u,x_i}(F_{x_i})}{\varepsilon}\right]du\\\\ =&\frac{1}{B-\alpha_1}\int_{\alpha_1}^{B}\left[\frac{d}{d\varepsilon}\underline{\VaR}_{u,x_i}(F_{\varepsilon,x_i})\right]_{\varepsilon=0}du\\\\
=&\begin{dcases} 
     \frac{1}{B-\alpha_1}\int_{\alpha_1}^{B}-\frac{A-u}{f_{x_i}\left[\underline{\VaR}_{u,x_i}(\VX)\right]A}du,&\ \   z<\underline{\VaR}_{\alpha,x_i}(\VX),\\
     \frac{1}{B-\alpha_1}\left\{\int_{\alpha_1}^{F(x_i,z)}\frac{u}{f_{x_i}\left[\underline{\VaR}_{u,x_i}(\VX)\right]A}du\right.\\\\\left.+\int_{F(x_i,z)}^{B}-\frac{A-u}{f_{x_i}\left[\underline{\VaR}_{u,x_i}(\VX)\right]A}du\right\}, &\ \  \underline{\VaR}_{\alpha,x_i}(\VX)\leq z\leq \VaR_{\alpha_2}(X_j),\\
     \frac{1}{B-\alpha_1}\int_{\alpha_1}^{B}\frac{u}{f_{x_i}\left[\underline{\VaR}_{u,x_i}(\VX)\right]A}du, &\ \  z>VaR_{\alpha_2}(X_j).\\
  \end{dcases}\\\\
  =&S'(z)-\underline{\RVaR}_{\alpha_1,\alpha_2,x_i}(\VX),
\end{align*}
where 

 \begin{align*}
 S'(z)=&
 \begin{dcases} 
     \frac{(A-\alpha_1)\underline{\VaR}_{\alpha_1,x_i}(\VX)-(A-B)\VaR_{\alpha_2}(X_j)}{B-\alpha_1},&\ \   z<\underline{\VaR}_{\alpha_1,x_i}(\VX),\\
     \frac{zA-\alpha_1\underline{\VaR}_{\alpha_1,x_i}(\VX)-(A-B)\VaR_{\alpha_2}(X_j)}{B-\alpha_1}, &\ \  \underline{\VaR}_{\alpha_1,x_i}(\VX)\leq z\leq \VaR_{\alpha_2}(X_j),\\
     \frac{B \VaR_{\alpha_2}(X_j)-\alpha_1\underline{\VaR}_{\alpha_1,x_i}(\VX)}{B-\alpha_1}, &\ \  z>VaR_{\alpha_2}(X_j).\\
  \end{dcases}
\end{align*}

Furthermore, the sensitivity function of $\underline{\TVaR}_{\alpha,x_i}(\VX)$ can be obtained when $B=A$.

Then,
\begin{align*}
S(z)&=\begin{dcases} 
     \underline{\VaR}_{\alpha,x_i}(\VX)-\underline{\TVaR}_{\alpha,x_i}(\VX),&\ \   z<\underline{\VaR}_{\alpha,x_i}(\VX),\\
     \frac{zA-\alpha\underline{\VaR}_{\alpha,x_i}(\VX)}{A-\alpha}-\underline{\TVaR}_{\alpha,x_i}(\VX), &\ \  z\geq\underline{\VaR}_{\alpha,x_i}(\VX).\\
  \end{dcases}
\end{align*}
Obviously, it is linear in $z$, which implies that $\underline{\TVaR}_{\alpha,x_i}(\VX)$ is not a robust statistic. This also coincides with univariate TVaR.
\end{proof}

\subsection*{Proposition \ref{Proposition:UpperVaRRobust}}

\begin{proof}
Let $F_ {\bar{x}_i}(x_j)=\Pr(X_j\leq x_j|X_i\geq x_i)$ be the conditional distribution of $X_j$ given $X_i\geq x_i$, $i,j=1,2$. For any fixed $x_i$ and $\varepsilon\in[0,1)$, set $F_{\varepsilon,\bar{x}_i}(x_j)=\varepsilon\delta_{z}+(1-\varepsilon)F_{\bar{x}_i}(x_j)$. $F_{\varepsilon,\bar{x}_i}$ is differentiable at any $x_j\neq z$ with $F_{\varepsilon,\bar{x}_i}^{'}(x_j)=(1-\varepsilon)f_{\bar{x}_i}(x_j)>0$ and has a jump at the point $x_j=z$.\\
Then, given that $\overline{\VaR}_{\alpha,x_i}(\VX)=\VaR_{\frac{\alpha-F_{X_i}(x_i)}{1-F_{X_i}(x_i)}}(X_j|X_i\geq x_i),$
we have
\begin{align*}
\overline{\VaR}_{\alpha,x_i}(F_{\varepsilon,\bar{x}_i})&=F^{-1}_{\varepsilon,\bar{x}_i}\left(\frac{\alpha-F_{X_i}(x_i)}{1-F_{X_i}(x_i)}\right)\\
&=\begin{dcases} 
     F^{-1}_{\bar{x}_i}\left(\frac{\alpha-F_{X_i}(x_i)}{(1-\varepsilon)(1-F_{X_i}(x_i))}\right),&\ \   \frac{\alpha-F_{X_i}(x_i)}{1-F_{X_i}(x_i)}<(1-\varepsilon)F_{\bar{x}_i}(z),\\
     F^{-1}_{\bar{x}_i}\left(\frac{\frac{\alpha-F_{X_i}(x_i)}{1-F_{X_i}(x_i)}-\varepsilon}{1-\varepsilon}\right), &\ \  \frac{\alpha-F_{X_i}(x_i)}{1-F_{X_i}(x_i)}\geq(1-\varepsilon)F_{\bar{x}_i}(z)+\varepsilon,\\
      z, &\ \  \text{otherwise}.\\
  \end{dcases}
\end{align*}
Hence, the sensitivity function of $\overline{\VaR}_{\alpha,x_i}(\VX)$ can be obtained by

\begin{align*}
S(z) &=\lim_{\varepsilon\rightarrow0^{+}}\frac{\overline{\VaR}_{\alpha,x_i}(F_{\varepsilon,\bar{x}_i})-\overline{\VaR}_{\alpha,x_i}(F_{\bar{x}_i})}{\varepsilon}\\ &=\left[\frac{d}{d\varepsilon}\overline{\VaR}_{\alpha,x_i}(F_{\varepsilon,\bar{x}_i})\right]_{\varepsilon=0}\\
&=\begin{dcases} 
    -\frac{1-\alpha}{f_{\bar{x}_i}\left[\overline{\VaR}_{\alpha,x_i}(\VX)\right](1-F_{X_i}(x_i))},&\ \   z<\overline{\VaR}_{\alpha,x_i}(\VX),\\
     \frac{\alpha-F_{X_i}(x_i)}{f_{\bar{x}_i}\left[\overline{\VaR}_{\alpha,x_i}(\VX)\right](1-F_{X_i}(x_i))}, &\ \  z>\overline{\VaR}_{\alpha,x_i}(\VX),\\
      0, &\ \  z=\overline{\VaR}_{\alpha,x_i}(\VX).\\
  \end{dcases}
\end{align*}
As $\underline{\VaR}_{\alpha,x_i}(\VX)$, $\overline{\VaR}_{\alpha,x_i}(\VX)$ also has a bounded sensitivity function, meaning it is also robust. And differences in results is because that bivariate lower and upper orthat RVaR are evaluated using cdf and survival function, respectively.\\
\end{proof}

\subsection*{Proposition \ref{Proposition:RVaRUpperRobust}}

\begin{proof}
 Let $A=F_{X_i}(x_i)$ and $C=1-\bar{F}(x_i,VaR_{\alpha_1}(X_j))$. Then the sensitivity function of $$\overline{\RVaR}_{\alpha_1,\alpha_2,x_i}(\VX)=\frac{1}{\alpha_2-C}\int_{C}^{\alpha_2}\overline{\VaR}_{v,x_i}(\VX)dv,$$ is given by
\begin{align*}
S(z) =&\frac{1}{\alpha_2-C}\int_{C}^{\alpha_2}\left[\lim_{\varepsilon\rightarrow0^{+}}\frac{\overline{\VaR}_{v,x_i}(F_{\varepsilon,\bar{x}_i})-\overline{\VaR}_{v,x_i}(F_{\bar{x}_i})}{\varepsilon}\right]dv\\\\ =&\frac{1}{\alpha_2-C}\int_{C}^{\alpha_2}\left[\frac{d}{d\varepsilon}\overline{\VaR}_{v,x_i}(F_{\varepsilon,\bar{x}_i})\right]_{\varepsilon=0}dv\\\\
&=\begin{dcases} 
    \frac{1}{\alpha_2-C}\int_{C}^{\alpha_2}-\frac{1-v}{f_{\bar{x}_i}\left[\overline{\VaR}_{v,x_i}(\VX)\right](1-A)}dv,&\ \   z<\VaR_{\alpha_1}(X_j),\\
     \frac{1}{\alpha_2-C}\left\{\int_{C}^{1-\bar{F}(x_i,z)}\frac{v-A}{f_{\bar{x}_i}\left[\overline{\VaR}_{v,x_i}(\VX)\right](1-A)}dv\right.\\\left.+\int_{1-\bar{F}(x_i,z)}^{\alpha_2}-\frac{1-v}{f_{\bar{x}_i}\left[\overline{\VaR}_{v,x_i}(\VX)\right](1-A)}dv\right\}, &\ \  \VaR_{\alpha_1}(X_j)\leq z\leq \overline{\VaR}_{\alpha_2,x_i}(\VX),\\
      \frac{1}{\alpha_2-C}\int_{C}^{\alpha_2}\frac{v-A}{f_{\bar{x}_i}\left[\overline{\VaR}_{v,x_i}(\VX)\right](1-A)}dv, &\ \  z>\overline{\VaR}_{\alpha_2,x_i}(\VX).\\
  \end{dcases}\\\\
  &=S'(z)-\overline{\RVaR}_{\alpha_1,\alpha_2,x_i}(\VX),
\end{align*}

where

$ S'(z)=$
 \begin{align*}
\begin{dcases} 
     \frac{(1-C)\VaR_{\alpha_1}(X_j)-(1-\alpha_2)\overline{\VaR}_{\alpha_2,x_i}(\VX)}{\alpha_2-C},&\ \   z<\VaR_{\alpha_1}(X_j),\\
     \frac{z(1-A)-(C-A) \VaR_{\alpha_1}(X_j)-(1-\alpha_2)\overline{\VaR}_{\alpha_2,x_i}(\VX)}{\alpha_2-C}, &\ \  \VaR_{\alpha_1}(X_j)\leq z\leq \overline{\VaR}_{\alpha_2,x_i}(\VX),\\
     \frac{(\alpha_2-A) \overline{\VaR}_{\alpha_2,x_i}(\VX)-(C-A) \VaR_{\alpha_1}(X_j)}{\alpha_2-C}, &\ \  z>\overline{\VaR}_{\alpha_2,x_i}(\VX).\\
  \end{dcases}
\end{align*}

Furthermore, the sensitivity function of $\overline{\TVaR}_{\alpha,x_i}(\VX)$ can be obtained, when $\beta=\alpha$ and $\alpha_2=1$. Then,

\begin{align*}
S(z)
&=\begin{dcases} 
    \overline{\VaR}_{\alpha,x_i}(\VX)-\overline{\TVaR}_{\alpha,x_i}(\VX),&\ \   z<\overline{\VaR}_{\alpha,x_i}(\VX),\\
     \frac{z(1-A)-(\alpha-A)\overline{\VaR}_{\alpha,x_i}(\VX)}{1-\alpha}-\overline{\TVaR}_{\alpha,x_i}(\VX), &\ \  z\geq\overline{\VaR}_{\alpha,x_i}(\VX).\\
  \end{dcases}
\end{align*}

Because of their analogous definitions, the sensitivity function of $\overline{\TVaR}_{\alpha,x_i}(\VX)$ is similar to the one of $\underline{\TVaR}_{\alpha,x_i}(\VX)$. Consequently, $\overline{\TVaR}_{\alpha,x_i}(\VX)$ is not robust.
\end{proof}

\subsection*{Proposition \ref{Proposition:Empirical}}
\begin{proof}
According to Theorem 2.1 by \citet{beck2015multivariate}, we have
$$\underline{\VaR}^n_{u,x_{i}}(\VX)\underset{n\longrightarrow\infty}{\overset{wp1}{\longrightarrow}}\underline{\VaR}_{u,x_{i}}(\VX).$$
for any $u\in(0,1)$. From the assumption, one has
$$F_{n,i}\underset{n\longrightarrow\infty}{\overset{wp1}{\longrightarrow}}F_i.$$ Then,

\begin{align*}
\textbf{1}_{\left[\alpha_1, F_n(x_{i},\VaR_{\alpha_2}(X_j))\right]}(u)=&
    \begin{dcases}
      1, & u\in\left[\alpha_1, F_n(x_{i},\VaR_{\alpha_2}(X_j))\right] \\
      0, & \text{otherwise}
    \end{dcases}\\
\underset{n\longrightarrow\infty}{\overset{wp1}{\longrightarrow}}&
    \begin{dcases}
      1, & u\in\left[\alpha_1, F(x_{i},\VaR_{\alpha_2}(X_j))\right] \\
      0, & \text{otherwise}
    \end{dcases}\\
=&\textbf{1}_{\left[\alpha_1, F(x_{i},\VaR_{\alpha_2}(X_j))\right]}(u).
\end{align*}
As a result, it can be seen that\\
$$\frac{\underline{\VaR}^n_{u,x_{i}}(\VX)\textbf{1}_{\left[\alpha_1, F_n(x_{i},\VaR_{\alpha_2}(X_j))\right]}(u)}{F_n(x_{i},\VaR_{\alpha_2}(X_j))-\alpha_1}\underset{n\longrightarrow\infty}{\overset{wp1}{\longrightarrow}}\frac{\underline{\VaR}_{u,x_{i}}(\VX)\textbf{1}_{\left[\alpha_1, F(x_{i},\VaR_{\alpha_2}(X_j))\right]}(u)}{F(x_{i},\VaR_{\alpha_2}(X_j))-\alpha_1}.$$

Therefore, by the dominated convergence theorem,\\
\begin{align*}
\lim_{n\rightarrow\infty}\underline{\RVaR}^n_{\alpha_1,\alpha_2,x_i}(\VX)&=\lim_{n\rightarrow\infty}\int \frac{\underline{\VaR}^n_{u,x_{i}}(\VX)\textbf{1}_{\left[\alpha_1, F_n(x_{i},\VaR_{\alpha_2}(X_j))\right]}(u)}{F_n(x_{i},\VaR_{\alpha_2}(X_j))-\alpha_1}du\\
&= \int\frac{\underline{\VaR}_{u,x_{i}}(\VX)\textbf{1}_{\left[\alpha_1, F(x_{i},\VaR_{\alpha_2}(X_j))\right]}(u)}{F(x_{i},\VaR_{\alpha_2}(X_j))-\alpha_1}du\\
&=\frac{\int_{\alpha_1}^{F(x_{i},\VaR_{\alpha_2}(X_j))}\underline{\VaR}_{u,x_{i}}(\VX)du}{F(x_{i},\VaR_{\alpha_2}(X_j))-\alpha_1}  \\
&=\underline{\RVaR}_{\alpha_1,\alpha_2,x_i}(\VX).
\end{align*}

Note that the consistency of upper orthant RVaR could be proved in the same way.
\end{proof}

\bibliographystyle{apalike}
\bibliography{mybib.bib}

\end{document}